\documentclass[journal,comsoc]{IEEEtran}
\usepackage{amsfonts}
\usepackage{amssymb}
\usepackage{eurosym}
\usepackage{cite}
\usepackage{graphicx}
\usepackage{epstopdf}
\usepackage{amsmath}
\usepackage{tikz,lipsum}
\usepackage{caption}
\usepackage[T1]{fontenc}
\usepackage{amsthm}
\usepackage{mathrsfs}
\usepackage{color}
\usepackage{stfloats}
\usepackage{xcolor, etoolbox}
\usepackage{subcaption}
\usepackage{comment}
\usepackage{algcompatible}
\usepackage{authblk}
\usepackage[ruled,linesnumbered]{algorithm2e}

\IEEEoverridecommandlockouts
\newcounter{mytempeqncnt}
\newtheorem{remark}{Remark}
\newtheorem{theorem}{Theorem}

\begin{document}

\title{On the Secrecy-Sensing Optimization of RIS-assisted Full-Duplex
Integrated Sensing and Communication Network}
\author{Elmehdi Illi, \IEEEmembership{Member, IEEE}, Ahmad Bazzi, %
\IEEEmembership{Member, IEEE}, Marwa Qaraqe,
\IEEEmembership{Senior
Member, IEEE}, and Ali Ghrayeb, \IEEEmembership{Fellow, IEEE} \thanks{%
E. Illi, M. Qaraqe, and A. Ghrayeb are with the College of Science and
Engineering, Hamad Bin Khalifa University, Doha, Qatar. (e-mails:
elmehdi.illi@ieee.org, aghrayeb@hbku.edu.qa, mqaraqe@hbku.edu.qa.)} \thanks{%
A. Bazzi is with the Engineering Division, New York University (NYU) Abu
Dhabi, Abu Dhabi, United Arab Emirates, and NYU WIRELESS, NYU Tandon School
of Engineering, Brooklyn, 11201, NY, USA. (e-mail: ahmad.bazzi@nyu.edu)}
\thanks{%
This work has been partially supported by Tamkeen and the Center for
Cybersecurity under the NYU Abu Dhabi Research Institute Award G$1104$.} }
\maketitle

\begin{abstract}
Integrated sensing and communication (ISAC) has recently emerged as a viable technique for establishing sensing and communication using the same
resources. Nonetheless, the operation of ISAC networks is often challenged by the absence of a direct link between the sensing node and the targets, and by the risk of disclosing confidential data to malicious targets
when using the same signal for both tasks. In this paper, a robust reconfigurable intelligent surface (RIS)-aided scheme for securing a
full-duplex (FD) ISAC network is proposed. The considered network
consists of uplink and downlink users served in FD through a multi-antenna
dual-functional radar communication base station (BS), which employs
co-located multi-antenna communication-radar arrays to detect multiple
malicious targets while preserving communication secrecy in their presence. Additionally, the BS utilizes an optimized
artificial noise (AN) that serves to disrupt the malicious targets' reception and increase the sensing power. By optimally
designing the RIS phase shifts, transmit beamforming, AN covariance, and
uplink users' transmit power and combining vectors using an alternating
optimization-based algorithm, the network's sensing performance is maximized
under secrecy and total power constraints. Numerical results present the proposed scheme's efficacy, particularly when a direct
link between the BS and the various nodes/targets is absent.
\end{abstract}

\begin{IEEEkeywords}
Artificial noise, dual-functional radar communication, eavesdropping, full-duplex communications, integrated sensing and communication (ISAC), physical-layer security, and reconfigurable intelligent surfaces (RIS).
\end{IEEEkeywords}

\section{Introduction}

\subsection{Background}

The vision for the sixth-generation (6G) of wireless networks envisions to
bolster the conventional key performance indicators (KPIs) established in 5G
and earlier generations, such as data rate and coverage, in addition to
sensing accuracy as a critical KPI \cite{survey6g}. In practice, sensing and
communication tasks are allocated different chunks from the radio spectrum
in order to avoid mutual interference. For instance, several designs have
been proposed to allow radar sensing and communication to coexist
concurrently by using different spectrum portions and dedicated hardware for
each of the two tasks \cite{refmag}. Such a design is generally power-,
spectrum-, and hardware-inefficient as separate hardware, resources (e.g.,
frequencies, time slots), and power sources are utilized to serve both tasks.

Recently, the integrated sensing and communication (ISAC) concept has
attracted a significant amount of interest, where the objective is to enable
both sensing and communication using the same hardware and spectrum
resources \cite{refmag2}. Thus, ISAC can pave the way for integrating both
tasks within the same transceiver in a more power- and spectrum-efficient
way. Nonetheless, despite the aforementioned ISAC advantages, its operation
faces some challenges. For instance, the use of the same signal/waveform for
both tasks puts communication confidentiality at stake as sensed targets can
be malicious entities that might illegitimately decode the ISAC signal. Additionally, the absence of a direct link between the sensing ISAC transceiver (e.g., radar base station (BS)) and the targets makes object detection and localization extremely challenging. Therefore, it is of paramount importance to design robust and \textit{secure} ISAC networks,
maintaining an acceptable secrecy-sensing trade-off.

Physical layer security (PLS) has attained significant attention over the
past decades as means for establishing theoretically-secure transmission.
PLS leverages physical-layer parameters along with wireless transmission
techniques to safeguard data confidentiality by increasing the received
confidential signal power at the legitimate receivers, while decreasing the
malicious users' (i.e., eavesdroppers) signal-to-noise ratios (SNRs),
yielding a degraded decoding capability at the latter. Unlike conventional
complex cryptographic-based security techniques that are
computationally-secure, PLS techniques can achieve theoretically-secure
transmissions irrespective of the adversaries computational power \cite%
{plssurvey}.

From another front, the emerging reconfigurable intelligent surfaces (RIS)
technology contributed to the improvement of wireless network's power- and
spectrum-efficiencies, thanks to its controllable reflection property \cite%
{direnzo}. Such engineered man-made reflecting surfaces are designed to
reflect the impinging electromagnetic waves with proper phase shifts to
electronically steer it into the intended receiver/target with a high power.
Thus, RIS has been notably incorporated in designing information-theoretic secure wireless networks by designing optimal RIS phase shifts either increasing the signal power at the legitimate receiver or degrading it at
the eavesdropper \cite{plsris1,plsris2}. From an ISAC perspective, RIS
involvement substantially benefits the sensing aspect by creating a virtual
line-of-sight (LoS) for targets when no LoS (direct link) to the radar
transceiver is available \cite{risisac}.

\subsection{Related Work}

The corresponding literature includes several work proposing robust designs
for secure ISAC networks, particularly for dual-functional radar
communication (DFRC)-based ISAC schemes. In these work, a common assumption
is to consider the sensed target as a malicious node. For instance, the work
in \cite{lit13} proposes a secure DFRC ISAC scheme by optimizing the
transmit legitimate and sensing signals covariance matrices in the presence
of multiple targets. Similarly, the scheme in \cite{lit14} aims at
maximizing the sensing performance subject to secrecy and power constraints,
assuming an imperfect channel state information (CSI) for legitimate users
and eavesdroppers. The authors of \cite{lit15} propose a secure DFRC ISAC
network design, under the consideration of an uncertain malicious target
location. The proposed design focuses on maximizing the worst-case secrecy
performance with respect to all possible eavesdropper's directions subject
to a maximal posterior Cram\'er-Rao bound, considered as a sensing
constraint. Furthermore, the work in \cite{lit16} investigates the
secrecy-sensing performance optimization of a non-orthogonal multiple access
(NOMA)-based multi-user ISAC network. The transmit signal beamforming and
artificial noise (AN) covariance matrices are optimized for maximizing the
secrecy rate of the network subject to a minimal echo SNR level. In \cite%
{refbench1,lit17,lit18,lit11}, the authors deal with the secrecy-sensing
optimization of a DFRC-based ISAC scheme considering unknown initial
eavesdroppers' locations. An initial sensing procedure is performed over a
large angular interval to obtain the eavesdroppers' locations, which then
serves as a basis for establishing an optimized beamforming approach. In
\cite{bazzi}, the authors optimize the total power consumption of a
full-duplex (FD) DFRC-ISAC network subject to secrecy and sensing
constraints. By the use of an optimized AN and legitimate signal
beamforming, uplink (UL) and downlink (DL) data confidentiality is preserved
while maintaining an acceptable sensing performance.

From another front, the incorporation of RIS to enhance the secrecy and
sensing performance of ISAC networks stimulated several research efforts.
For instance, the authors in \cite{lit1, lit2, lit3} focus on optimizing the
RIS phase shift, transmit beamforming, and AN covariance matrix to maximize
the secrecy performance of a DFRC-ISAC network, with either
uncertain/unavailable eavesdroppers CSI. The designed schemes aim at either
maximizing the secrecy rate subject to a target sensing requirement, or
vice-versa. An enhanced scheme is analyzed in \cite{lit9} considering an
aerial RIS, where this latter's position is also considered in the control
variables for the same secrecy and sensing requirements. The authors in \cite%
{lit5} propose a reinforcement learning (RL)-based scheme for optimizing
RIS\ and BS beamforming in dynamic ISAC\ networks. Jiang et al. in \cite%
{lit4} extend the secure RIS-aided ISAC\ system design and optimization for
a NOMA multi-user network, taking into account optimizing the same
aforementioned control variables, while \cite{lit6,lit7,lit8} consider the
use of an active RIS for improving ISAC\ networks' secrecy-sensing
trade-off. The concept of movable antennas is introduced in \cite{lit10} to
increase the degrees of freedom in secure RIS-aided ISAC\ networks.

\subsection{Paper Contributions}

While the aforementioned work, particularly the ones involving the use of
RIS contribute notably to the secrecy and sensing optimization of ISAC
networks, they are restricted mainly to half-duplex transmission. The
adoption of an FD transmission can double the effective rate by allowing
both UL and DL transmission using the same band \cite{fullduplex}. In the
context of secure ISAC, an FD DFRC BS aims at serving both UL and DL users
while sensing several malicious targets and preserving data leakage to them
as much as possible. It is worth recalling that the only two works on
\textbf{secure FD-ISAC} networks are \cite{lit19,bazzi}. In \cite{bazzi},
the proposed scheme consists of an FD ISAC network with multiple UL and DL
users and various malicious targets, whereas the setup in \cite{lit19}
assumes the presence of a single UL and a single malicious target only.
Although the former scheme (which generalizes the latter) considers multiple
malicious targets and UL/DL users, when direct links to these users and
targets are fully blocked, it becomes impossible to establish secure
transmissions or reliably sense and detect the targets.

Thus, motivated by the above, we aim in this paper to design a RIS--aided
scheme for enhancing the secrecy-sensing trade-off in an FD ISAC network.
Unlike the aforementioned RIS-aided ISAC schemes \cite%
{lit1,lit2,lit3,lit4,lit5,lit6,lit7,lit8,lit9,lit10} focusing only on DL
transmission and sensing, the proposed work involves UL transmissions as
well, whereby the UL users' powers and receive combining vectors, RIS phase
shifts, transmit beamforming and AN design are optimized to maximize the
network's sensing performance subject to secrecy and total power
constraints. The proposed scheme aims at circumventing the direct link
absence for an FD-DFRC-ISAC network by enabling RIS-aided reflection. The
main contributions of this paper can be summarized as follows:

\begin{itemize}
\item A RIS-aided FD-DFRC ISAC system, considering multiple UL and DL users
and various malicious targets is considered and modeled in terms of its main
system parameters. The malicious targets aim at overhearing the legitimate
data in both UL and DL.

\item Due to the high coupling between (i) the receive BS's UL combiners,
transmit beamforming, AN covariance matrix, UL users' powers, and (ii) the
RIS phase shifts, a robust alternating optimization (AO)-based algorithm is
proposed for jointly optimizing the aforementioned variables alternatively.
The proposed approach achieves convergence after a few iterations.

\item By adopting a successive convex approximation (SCA) technique and a
proposed iterative approximated optimization, the non-convex UL legitimate
signal-to-interference-and-noise ratio (SINR) constraints in the inner
subproblems of the AO-based framework are convexified.

\item Elaborative numerical evaluation is performed for the proposed
framework showing its effectiveness in maximizing the sensing performance,
while always fulfilling the secrecy requirements in both UL and DL. The
proposed scheme also outperforms its benchmark RIS-less FD-ISAC one when
direct link communications are absent and when the eavesdroppers are aligned
with the legitimate users with respect to the BS.

\end{itemize}

\subsection{Paper Organization}

The remainder of this paper is organized as follows: Section II presents the
analyzed system and channel models as well as the considered secrecy and
sensing metrics. In Section III, the considered secrecy-sensing optimization
problem is provided and the adopted algorithmic approach for solving it is
detailed, while Section IV provides illustrative numerical results for the
proposed scheme. Finally, Section V concludes the paper.

\subsection{Notations}

$i$ indicates the imaginary unit number $(i^2=-1)$, upper-case and
lower-case bold symbols (e.g., $\mathbf{H}$ and $\mathbf{h}$)\ indicate
matrices and vectors, respectively, $\otimes $ is the Kronecker product, $%
\mathbb{E}\left[ \mathbf{.}\right] $ refers to the expectation operator, $%
\mathbf{I}_{L}$ is the $L\times L$ identity matrix, $\mathrm{Tr}\left[
\mathbf{H}\right] $ refers to the trace of a matrix $\mathbf{H}$, $\mathrm{%
diag}\left( \mathbf{h}\right) $ transforms a vector $\mathbf{h}$ into a
diagonal matrix $\mathbf{H}$ whose diagonal entries are the elements of $%
\mathbf{h}$. Also, the superscripts $\left( .\right) ^{T}$ and $\left(
.\right) ^{H}$ refer to the transpose and conjugate transpose of a matrix or
vector, respectively, while $\left[ \mathbf{H}\right] _{m,n}$ refers to $%
\mathbf{H}$'s entry at its $m$th row and $n$th column. $\mathbf{0}_{M \times N}$ and $\mathbf{1}_{M \times N}$ stand for $M$-by-$N$ matrices of zeros and ones, respectively, and $\mathcal{CN}(\mu,\sigma^2)$ denote the complex Gaussian distributed with mean $\mu$ and variance $\sigma^2$. Also, $\left[\mathbf{D%
}\right]_{m,:}$ and $\left[\mathbf{D%
}\right]_{:,n}$ define, respectively, the $m$th row and $n$th column of a matrix $\mathbf{D}$, and $%
\angle.$ and $\left| .\right|$ are, respectively, the argument and the absolute value of a complex number, and $\left[ x\right]^+ \triangleq \max(0,x)$. Lastly, $\mathrm{maxeigv}%
\left( \mathbf{H}\right) $ returns the eigenvector of $\mathbf{H}$
corresponding to the maximal eigenvalue of the same matrix.

\section{System and Channel Models}

\begin{figure}[tbp]
\begin{center}
\includegraphics[scale=.25]{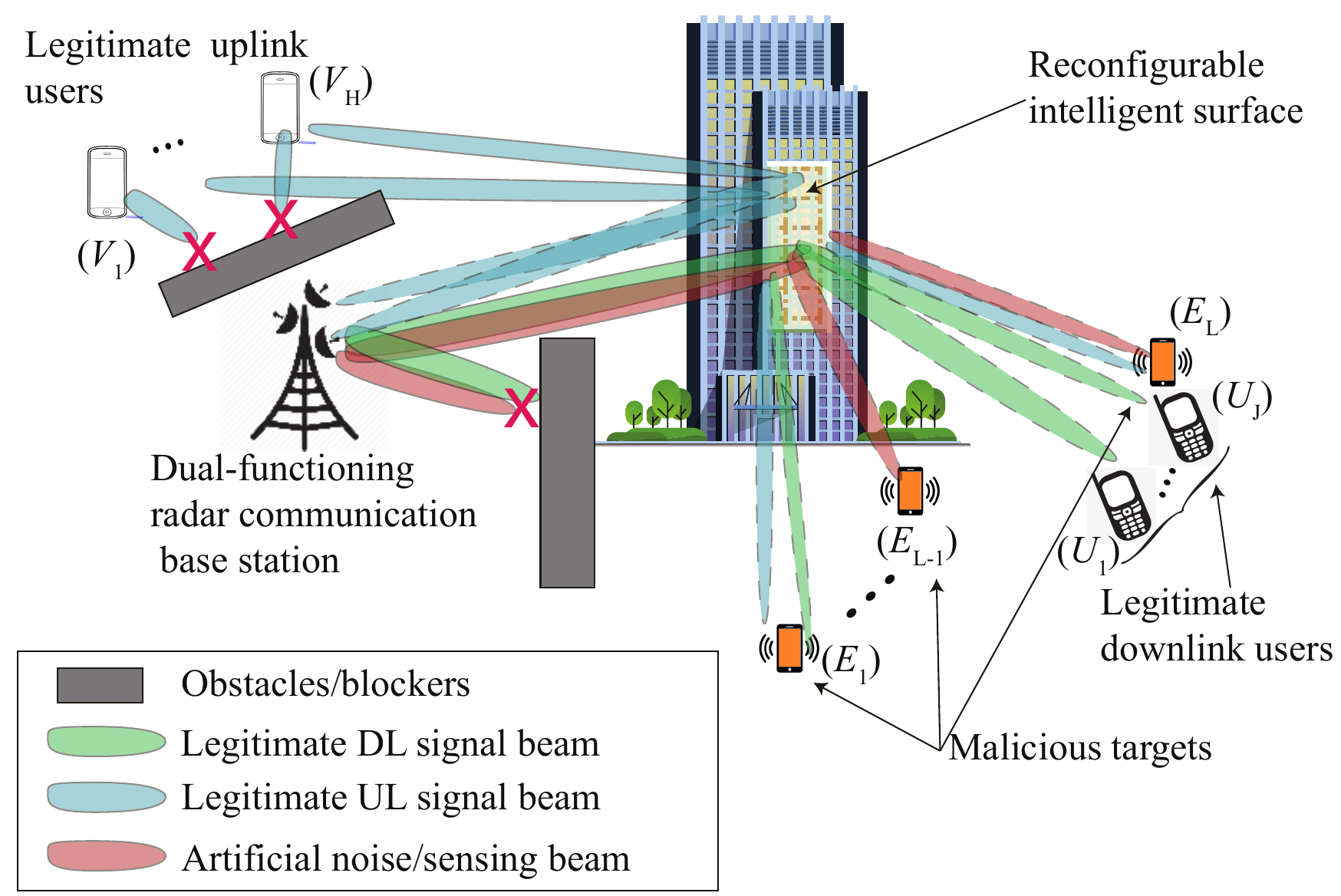}
\end{center}
\caption{{Considered ISAC\ network mode:\ Dashed-lined beams are the ones
reflected by the RIS.}}
\label{sysmod}
\end{figure}

\subsection{Communication and Eavesdropping Models}

We consider an FD DFRC RIS-assisted wireless network, as shown in Fig. \ref%
{sysmod}, where a FD-DFRC BS, labeled $B$, and equipped with $N_{t}$
transmit antennas and $N_{r}$ receive ones, aims to communicate with a set
of $J$ single-antenna DL receivers, denoted by $\left\{ U_{j}\right\}
_{j=1}^J$ while simultaneously receiving from $K$ UL users $\left\{
V_{k}\right\} _{k=1}^K$. For this, the BS\ exploits its transmit and
receive antenna array to generate several DL\ beams to the DL\ users while
combining the received signals at its $N_{r}$ antennas from each UL\ user.
Due to the absence of a direct link between the BS\ and the various UL\ and
DL\ users\footnote{%
The direct link can be unavailable due to the presence of obstructing and absorbing
objects (e.g., walls, buildings) in the network with an excessively high
shadowing loss \cite{5G1,ETSI}. Therefore, in the considered network, the use of RIS is one
approach to enable UL and DL communications, in addition to sensing.}, an $N$%
-reflective elements (REs) RIS $(R)$ is incorporated, with square-shaped REs
of length $\Delta _{r}$ each, for creating a virtual LoS\ and enable
signal reflection to $\left\{ U_{j}\right\} _{j=1}^J$ \ (DL) and $B$
(UL). Furthermore, the BS\ leverages its collocated transmit and receive
antenna arrays to perform monostatic radar sensing for $L$ malicious targets
$\left\{ E_{l}\right\} _{l=1}^L$ supposed to be acting as potential
eavesdroppers. The latter nodes aim at eavesdropping the legitimate UL and
DL signals broadcasted by UL\ users and the FD\ BS. Similar to communication
links, the $B-E_{l}$ links are assumed to be fully blocked. Herein, the RIS\
helps in creating a signal reflection from $B$ to the various targets to
perform the sensing task. For the latter task, the FD\ BS\ leverages an AN\
signal designed for a dual-purpose, namely:

\begin{enumerate}
\item To act as a sensing signal and enable target detection by virtue of
the received signal echoes back to $B$. The designed sensing beams hit the
various malicious targets via RIS-enabled reflection and similarly propagate
backwards to $B$.

\item To safeguard the communication secrecy with respect to the malicious
targets' eavesdropping by degrading the legitimate signal decoding
performance at the eavesdroppers with AN\ injection.
\end{enumerate}

To this end, the received downlink signal at $U_{j}$ $\left( j=1,\ldots
,J\right) $ or $E_{l}$ $\left( l=1,\ldots ,L\right) $ from $B$ can be
written as
\begin{equation}
y_{Z}=\mathbf{h}_{BZ}\mathbf{s}+\sum\limits_{k=1}^{K}\sqrt{p_{k}}%
h_{V_{k}Z}v_{k}+n_{Z},Z\in \left\{ U_{j},E_{l}\right\} ,  \label{rxsig1}
\end{equation}%
where%
\begin{equation}
\mathbf{h}_{BZ}=\sqrt{\mathcal{L}_{BRZ}}\mathbf{h}_{RZ} \mathbf{\Phi} \mathbf{H}_{BR} \in \mathbb{C}^{1 \times N_t}
\label{hbrz}
\end{equation}%
is the cascaded $B-Z$ channel coefficient
\begin{equation}
\mathcal{L}_{BRZ}=\frac{G_{T,B}G_{R,Z}\Delta _{r}^{4}}{d_{BR}^{2}d_{RZ}^{2}%
\left( 4\pi \right) ^{2}}  \label{fsplris}
\end{equation}%
is the free-space path loss attenuation between the BS\ and $Z$ via $R$, $%
d_{XZ}$ is the $X$-$Z$ link's distance for any pair of nodes $X$ and $Z$ such that $XZ\in
\left\{ BR,RB,RU_{j},RE_{l},V_{k}R\right\}$\footnote{The distance between the BS\ and each user is significantly larger than the
inter-element spacing ($\Delta _{r}$ or $\Delta _{a}$) within the transmit
array or RIS, resulting in approximately the same distance between all
transmit/receive/reflecting elements to each receiving node.}, $G_{T,B}$ and
$G_{R,Z}$ are, respectively, the transmit antenna gain of $B$ and the
receive antenna gain of $Z$ \cite[Eqs. (9.33)-(9.35)]{emilbook}.
Additionally,
\begin{equation}
\mathbf{H}_{BR}=\sqrt{\frac{\kappa _{BR}}{\kappa _{BR}+1}}\mathbf{H}_{BR}^{%
\mathrm{(LoS)}}+\sqrt{\frac{1}{\kappa _{BR}+1}}\mathbf{H}_{BR}^{\mathrm{%
(NLoS)}}  \label{Hbr}
\end{equation}%
\begin{equation}
\mathbf{h}_{RZ}=\sqrt{\frac{\kappa_{RZ} }{\kappa _{RZ}+1}}\mathbf{h}_{RZ}^{%
\mathrm{(LoS)}}+\sqrt{\frac{1}{\kappa _{RZ}+1}}\mathbf{h}_{RZ}^{\mathrm{%
(NLoS)}},  \label{hrz}
\end{equation}%
are the $B$-$R$ and $R$-$Z$ links' channel matrix and vector, respectively,
with $\mathbf{H}_{BR}\in
\mathbb{C}
^{N\times N_{t}}$ and $\mathbf{h}_{RZ}\in
\mathbb{C}
^{1\times N}$, composed of a rank-one LoS\ components' matrix/vector, i.e.,%
\begin{equation}
\mathbf{H}_{BR}^{\mathrm{(LoS)}}=\mathbf{a}_{R}^{T}\left( \theta
_{RB},\varphi _{RB}\right) \mathbf{a}\left( \theta _{BR},\varphi
_{BR},\Delta _{a},N_{t}\right) \in \mathbb{C}^{N \times N_t},  \label{Hbrlos}
\end{equation}%
\begin{equation}
\mathbf{h}_{RZ}^{\mathrm{(LoS)}}=\mathbf{a}_{R}\left( \theta _{RZ},\varphi
_{RZ}\right) \in \mathbb{C}^{1 \times N} ,  \label{hrzlos}
\end{equation}%
where%
\begin{equation}
\mathbf{a}_{R}\left( \theta,\varphi\right) =\mathbf{a}\left(
0,\theta,\Delta _{r},N_{V}\right) \otimes \mathbf{a}\left( \theta
,\varphi,\Delta _{r},N_{H}\right) ,  \label{Ar}
\end{equation}%
and
\begin{equation}
\mathbf{a}\left( \theta ,\varphi ,\Delta ,M\right) \triangleq \left[
1,\ldots,e^{-i2\pi \Delta \left( M-1\right) \cos \left( \theta \right) \sin \left(
\varphi \right) /\lambda }
\right] \in \mathbb{C}^{1 \times M}.  \label{A}
\end{equation}%
In the above, $N_{H}$ and $N_{V}$ denote the number of REs in $R$
horizontally and vertically, respectively, suth that $N=N_{H}N_{V}$. Also, $%
\Delta _{a}$ is the inter-antenna spacing at $B$, and $\lambda $ is the
operating wavelength. For any pair of nodes $X$ and $Z$ such that $XZ\in
\left\{ BR,RB,RU_{j},RE_{l},V_{k}R\right\} ,$ $\varphi _{XZ}$ and $\theta
_{XZ}$ represent the azimuth and elevation angles between $X$ and $Z$,
respectively, defined with respect to the broadside direction of the transmitting, receiving, or reflecting array (i.e.,
perpendicular direction to $R$ or $B$'s array). In addition, $\kappa _{XZ}$
is the Rician K-factor for the $X$-$Z$ link, accounting for the ratio
between the LoS and non-LoS\ (NLoS) powers. Also, $\mathbf{H}_{BR}^{\mathrm{%
(NLoS)}}\in
\mathbb{C}
^{N\times N_{T}}$ and $\mathbf{h}_{RZ}^{\mathrm{(NLoS)}}\in
\mathbb{C}
^{1\times N}$ account for the NLoS\ components of the $B$-$R$ and $R$-$Z$
links' channels, respectively, whose elements are complex
Gaussian-distributed with zero mean and covariance matrix $\mathbb{E}\left[ %
\left[ \mathbf{H}_{BR}^{\mathrm{(NLoS)}}\right] _{:,m}\left( \left[ \mathbf{H%
}_{BR}^{\mathrm{(NLoS)}}\right] _{:,m}\right) ^{H}\right] =\mathbb{E}\left[
\mathbf{h}_{RZ}^{\mathrm{(NLoS)}}\left( \mathbf{h}_{RZ}^{\mathrm{(NLoS)}%
}\right) ^{H}\right] =\mathbf{I}_{N}$ $(\forall m)$. Also, $\Delta _{a}$ is
the BS\ array's inter-element spacing, $\mathbf{\Phi }=\mathrm{diag}\left(
\mathbf{q}\right) \in
\mathbb{C}
^{N\times N}$ represents the RIS\ reflection diagonal matrix with $\mathbf{%
q\triangleq }\left[ \phi _{1},\ldots ,\phi _{N}\right] $, $\phi
_{n}\triangleq e^{i\theta _{n}}$ and $\theta _{n}$ is the $n$th RE's phase
shift. Additionally,
\begin{equation}
\mathbf{s=Wx+z,}  \label{sig}
\end{equation}%
is the transmit beamformed signal, $\mathbf{W}\triangleq \left[ \mathbf{w}%
_{1},\ldots ,\mathbf{w}_{J}\right] \in
\mathbb{C}
^{N_{T}\times J}$ is the BS beamforming matrix to the $J$ DL\ users, $%
\mathbf{x\triangleq }\left[ x_{1},\ldots ,x_{J}\right] ^{T}\in
\mathbb{C}
^{J\times 1}$ is the unit-power downlink users' signal with an identity
covariance matrix, i.e., $\mathbf{C}_{x}=\mathbb{E}\left[ \mathbf{xx}^{H}%
\right] =\mathbf{I}_{J}$. Additionally, the AN$\ $signal$\ $vector $\mathbf{z%
}\in
\mathbb{C}
^{N_{t}\times 1}$ is superposed with the confidential signal to obfuscate it
and degrade its decoding performance at the eavesdroppers, which is a
complex Gaussian signal with zero mean and covariance matrix $\mathbf{C}_{z}=%
\mathbb{E}\left[ \mathbf{zz}^{H}\right] $. The transmit power of $V_{k}$ is $%
p_{k}$, $h_{V_{k}Z}$ is the $V_{k}-Z$ channel coefficient, arising from the
FD nature of the network, and $v_{k}$ is $V_{k}$'s transmit UL
signal with $\mathbb{E}\left[ v_{k}^{2}\right] =1$. $n_{Z}$ refers to the
additive white Gaussian noise (AWGN)\ at node $Z$ of zero mean and variance $%
\sigma _{n,Z}^{2}$. We further consider nodes of the same category to have
equal noise power, that is $\sigma _{n,U_{j}}^{2}=\sigma _{n,U}^{2}$ $%
\left( \forall j\right) $ and $\sigma _{n,E_{l}}^{2}=\sigma _{n,E}^{2}$ $%
\left( \forall l\right) $. Consequently,\ the received SINRs at $U_{j}$ for
the DL signal reception can be expressed from (\ref{rxsig1}) in the form of
matrix traces as%
\begin{equation}
\gamma _{U_{j}}=\frac{\mathrm{Tr}\left[ \mathbf{H}_{BU_{j}}\mathbf{W}_{j}%
\right] }{\sum\limits_{\substack{ j^{\prime }=1  \\ j^{\prime }\neq j}}^{J}%
\mathrm{Tr}\left[ \mathbf{H}_{BU_{j}}\mathbf{W}_{j^{\prime }}\right] +%
\mathrm{Tr}\left[ \mathbf{H}_{BU_{j}}\mathbf{C}_{z}\right]
+\sum\limits_{k=1}^{K}p_{k}\left\vert h_{V_{k}U_{j}}\right\vert ^{2}+\sigma
_{n,U}^{2}},  \label{snruj}
\end{equation}%
where $\mathbf{H}_{BZ}\triangleq \mathbf{h}_{BZ}^{H}\mathbf{h}_{BZ}$ for $%
Z\in \left\{ U_{j},E_{l}\right\} $ $\left( \forall j,l\right) $ and $\mathbf{%
W}_{j}\triangleq \mathbf{w}_{j}\mathbf{w}_{j}^{H}$ are hermitian positive
semidefinite matrices. Similarly, one can formulate the received SINR\ at
each eavesdropper $E_{l}$ for decoding $x_{j}$ $(j\in \left\{ 1,\ldots
,J\right\} )$ as%
\begin{equation}
\gamma _{E_{l},\mathrm{DL}}^{(j)}=\frac{\mathrm{Tr}\left[ \mathbf{H}_{BE_{l}}%
\mathbf{W}_{j}\right] }{\sum\limits_{\substack{ j^{\prime }=1  \\ j^{\prime
}\neq j}}^{J}\mathrm{Tr}\left[ \mathbf{H}_{BE_{l}}\mathbf{W}_{j^{\prime }}%
\right] +\mathrm{Tr}\left[ \mathbf{H}_{BE_{l}}\mathbf{C}_{z}\right]
+\sum\limits_{k=1}^{K}p_{k}\left\vert h_{V_{k}E_{l}}\right\vert ^{2}+\sigma
_{n,E}^{2}}.  \label{snrevedl}
\end{equation}

On the other hand, the UL received signals at $B$ and $E_{l}$ are
expressed, respectively as
\begin{equation}
\mathbf{y}_{B}=\sum\limits_{k=1}^{K}\sqrt{p_{k}}\mathbf{h}_{V_{k}B}v_{k}+%
\sqrt{\xi_{\mathrm{SI}}}\mathbf{H}_{BB}\mathbf{s+}\sum\limits_{l=1}^{L}\mathbf{H}_{BE_{l}B}\mathbf{s}%
+\mathbf{c}+\mathbf{n}_{B} \in \mathbb{C}^{N_r \times 1},  \label{rxsigbs}
\end{equation}%
and%
\begin{equation}
y_{E_{l}}=\sum\limits_{k=1}^{K}\sqrt{p_{k}}h_{V_{k}E_{l}}v_{k}+\mathbf{h}%
_{BE_{l}}\mathbf{s}+n_{E_{l}}.  \label{rxsigeveul}
\end{equation}%
Herein, $\mathbf{h}_{V_{k}B}=\sqrt{\mathcal{L}_{V_kRB}}\mathbf{H}_{RB} \mathbf{\Phi} \mathbf{h}_{V_kR}\in
\mathbb{C} ^{N_{r}\times 1}$ is the $V_{k}-B$ channel vector, with $\mathcal{L}_{V_kRB}$ defined from \eqref{fsplris} by swapping $B$ and $Z$ with $Z=V_k$, $\mathbf{H}_{RB} \in \mathbb{C}^{N_r \times N}$ is defined similarly to $\mathbf{H}_{BR}$ in \eqref{Hbr}, \eqref{Hbrlos} by substituting $N_t$ by $N_r$ and swapping the inner product order in \eqref{Hbrlos}, and \begin{equation}\mathbf{h}_{V_kR}=\sqrt{\frac{\kappa_{V_kR}}{\kappa_{V_kR}+1}}\mathbf{a}_R^T\left( \theta_{V_k R},\varphi_{V_k R}\right) + \sqrt{\frac{1}{\kappa_{V_kR}+1}} \mathbf{h}_{V_kR}^{(\mathrm{NLoS})}\in \mathbb{C}^{N \times 1}, \label{hvkr}
\end{equation}
with $\mathbf{h}_{V_kR}^{(\mathrm{NLoS})} \sim \mathcal{CN}(\mathbf{0}_{N \times 1},\mathbf{I}_N)$,
\begin{equation}
\mathbf{H}_{BB}=\mathbf{H}_{BB}^{\mathrm{(DL)}}+\sqrt{\mathcal{L}_{BRB}}%
\mathbf{H}_{RB}\Phi \mathbf{H}_{BR} \in \mathbb{C}^{N_r \times N_t} \label{channelsi}
\end{equation}%
is the self-interference channel matrix with $\left[ \mathbf{H}_{BB}^{%
\mathrm{(DL)}}\right] _{m,n}=e^{-j2\pi \Delta _{a}mn\sin \left( \varphi
_{0}\right) /\lambda }$ and $\mathcal{L}_{BRB}$ is the round-trip link's
path-loss computed from (\ref{fsplris}) with $Z=B$, $\mathbf{H}_{BE_{l}B}\in
\mathbb{C}
^{N_{r}\times N_{t}}$ is the round-trip link ($B$-$E_{l}$-$B$)'s channel
matrix. Also, $\mathbf{c}\in
\mathbb{C}
^{N_{r}\times 1}$ is the received radar clutter signal from undesired
reflections, and $\mathbf{n}_{B}$ is the zero-mean complex Gaussian AWGN\
vector at $B$, i.e., $\mathbf{n}_{B}\sim \mathcal{CN}\left( \mathbf{0}_{N_r \times 1},%
\sigma _{n,B}^{2}\mathbf{I}_{N_{r}}\right)$. Consequently, one can
deduce from \eqref{rxsigbs} that the SINR\ at the BS for decoding
$v_{k}$ is
\begin{equation}
\gamma _{B,k}=\frac{p_{h}\mathbf{r}_{k}^{H}\mathbf{h}_{V_{k}B}\mathbf{h}%
_{V_{k}B}^{H}\mathbf{r}_{k}}{\mathbf{r}_{k}^{H}\left(
\begin{array}{c}
\sum\limits_{\substack{ k^{\prime }=1 \\ k^{\prime }\neq k}}^{K}p_{k^{\prime
}}\mathbf{h}_{V_{k^{\prime }}B}\mathbf{h}_{V_{k^{\prime }}B}^{H}+\mathbf{H}%
_{BEB}\mathbf{C}_{s}\mathbf{H}_{BEB}^{H} \\
+\xi _{\mathrm{SI}}\mathbf{H}_{BB}\mathbf{C}_{s}\mathbf{H}_{BB}^{H}+\mathbf{R%
}_{c}+\sigma _{n,B}^{2}\mathbf{I}_{N_{r}}%
\end{array}%
\right) \mathbf{r}_{k}},  \label{snrbs}
\end{equation}%
where $\mathbf{r}_{k}\in \mathbb{C}^{N_{r}\times 1}$ is the BS's receive
combining vector for decoding $v_{k}$. Furthermore, $\mathbf{H}%
_{BEB}\in
\mathbb{C}
^{N_{r}\times N_{t}}$ is defined as the two-way equivalent channel matrix
for the $B$-$\left\{E_l  \right\}_{l=1}^L$-$B$ link, given as $\mathbf{H}_{BEB}=\widetilde{\mathbf{F}%
}_{EB}\mathrm{diag}\left( \boldsymbol{\rho }\right) \widetilde{\mathbf{F}}%
_{BE}$, with $\widetilde{\mathbf{F}}%
_{EB}\triangleq \mathbf{H}_{RB}\mathbf{\Phi} \mathbf{H}_{ER}\in \mathbb{C}%
^{N_{r}\times L}$ and $\widetilde{\mathbf{F}}_{BE}\triangleq \mathbf{H}%
_{RE}\mathbf{\Phi} \mathbf{H}_{BR}\in \mathbb{C}^{L\times N_{t}}$ representing the
path-loss-normalized cascaded channel matrices for the targets-BS\ and
BS-targets links, respectively, $\mathbf{H}_{ER}\in \mathbb{C}^{N\times L}$
and $\mathbf{H}_{RE}\in \mathbb{C}^{L\times N}$ are, respectively, the
Eves-RIS\ and RIS-Eves channel matrices, and $\boldsymbol{\rho }=\left[ \rho
_{1},\ldots ,\rho _{L}\right] $ with $\rho _{l}$ denoting the two-way
path-loss coefficient for the echo signal received from the $l$th target,
which can be derived in a similar way to \cite[Eqs. (8.71)]{emilbook} as
\begin{equation}
\rho _{l}=\left( \frac{\left( \Delta _{r}\lambda \right) ^{2}}{%
d_{BR}d_{RE_{l}}}\right) ^{2}\frac{G_{T,B}G_{R,B}\sigma _{RCS}^{(l)}}{\left(
4\pi \right) ^{5}},  \label{rhoo}
\end{equation}%
where $\sigma _{RCS}^{(l)}$ is the radar cross section (RCS) of the sensed
target in $\mathrm{m}^{2}$, measuring the effective area of the target when
facing the RIS. The proportionality of the above two-way path-loss term with
the respective RCS\ for each sensed target represents the target
detectability by the radar BS, where larger RCS\ values yield a higher
received echo power, resulting in enhanced target detection. Also, one can then
observe that $\mathbf{H}_{BE_{l}B}=\rho _{l}\widetilde{\mathbf{h}}_{E_{l}B}%
\widetilde{\mathbf{h}}_{BE_{l}}\in
\mathbb{C}
^{N_{r}\times N_{t}}$ ($\forall l$) with $\widetilde{\mathbf{h}}_{E_{l}B}=%
\left[ \widetilde{\mathbf{F}}_{EB}\right] _{:,l}$ and $\widetilde{\mathbf{h}}%
_{BE_{l}}=\left[ \widetilde{\mathbf{F}}_{BE}\right] _{l,:}$. Furthermore, $%
\xi _{\mathrm{SI}}$ is the residual self-interference power level at the BS
after cancelling it, and $\mathbf{R}_{c}\triangleq \mathbb{E}\left[ \mathbf{%
cc}^{H}\right] $ is the clutter covariance matrix. In a similar way, the
SINR\ at $E_{l}$ for decoding $V_{k}$'s signal is given as

\begin{equation}
\gamma _{E_{l},\mathrm{UL}}^{(k)}=\frac{p_{k}\left\vert
h_{V_{k}E_{l}}\right\vert ^{2}}{\sum\limits_{\substack{ k^{\prime }=1  \\ %
k^{\prime }\neq k}}^{K}p_{k^{\prime }}\left\vert h_{V_{k^{\prime
}}E_{l}}\right\vert ^{2}+\mathbf{h}_{BE_{l}}\mathbf{C}_{s}\mathbf{h}%
_{BE_{l}}^{H}+\sigma _{n,E}^{2}}.  \label{snreveul}
\end{equation}

\begin{remark}
\label{rkdirectlink} The RIS incorporation in the proposed scheme can be also
beneficial in scenarios when the direct link between the BS and the various
nodes is present. For instance, when the eavesdroppers lie in the same
azimuth angle as the legitimate DL users with respect to the BS (i.e.,
similar channel responses), it is not feasible to beamform the legitimate
signal beam and the AN from $B$ in two distinct directions. Thus, both types of beams
will be received by legitimate and illegitimate users, yielding an insecure
transmission. Therefore, the presence of RIS creates an indirect propagation
by reflection, enabling an angular separability of both types of beams.
\end{remark}

\subsection{Secrecy Evaluation Metrics}

From a PLS\ point of view, the secrecy capacity (SC)\ serves as a
representative metric of the network's theoretical secrecy limit, which
represents the maximal achievable rate ensuring both (i)\ a reliable signal
decoding at the intended (legitimate)\ receiver and (ii)\ a unit
equivocation rate at the eavesdropper \cite{plssurvey}. Such a metric is
defined as the difference between the channel capacities of the legitimate
and wiretap links as $C_{s}=\left[C_{L}-C_{E}\right]^+$ where $C_{L}\triangleq \log
_{2}\left( 1+\gamma _{L}\right) $ and $C_{E}\triangleq \log _{2}\left(
1+\gamma _{E}\right) $ are, respectively, the channel capacities of the
legitimate and the wiretap links, expressed in terms of their respective
SINRs. For the considered FD-DFRC-RIS\ scheme, due to the presence of
several eavesdroppers aiming to decode the signals of various UL and DL
users, we define the system's SC\ in UL and DL as \cite[Eqs. (21), (22)]%
{bazzi}%
\begin{equation}
C_{s,\mathrm{DL}}=\min_{\substack{ j=1,\ldots ,J  \\ l=1,\ldots ,L}}\left[
C_{U_{j}}-C_{E_{l}\mathrm{,DL}}^{(j)}\right] ^{+}  \label{Csdl}
\end{equation}%
and%
\begin{equation}
C_{s,\mathrm{UL}}=\min_{\substack{ k=1,\ldots ,K  \\ l=1,\ldots ,L}}\left[
C_{V_{k}}-C_{E_{l},\mathrm{UL}}^{(k)}\right] ^{+}  \label{Csul}
\end{equation}%
where $C_{U_{j}}=\log _{2}\left( 1+\gamma _{U_{j}}\right) $ and $%
C_{V_{k}}=\log _{2}\left( 1+\gamma _{B,k}\right) $ are the $j$th DL\ user's
and $k$th UL\ user's channel capacities, respectively, while%
\begin{equation}
C_{E_{l}\mathrm{,DL}}^{(j)}=\log _{2}\left( 1+\gamma _{E_{l},\mathrm{DL}%
}^{(j)}\right)  \label{ccdleve}
\end{equation}%
and%
\begin{equation}
C_{E_{l},\mathrm{UL}}^{(k)}=\log _{2}\left( 1+\gamma _{E_{l},\mathrm{UL}%
}^{(k)}\right)  \label{cculeve}
\end{equation}%
are the corresponding eavesdropping channel capacities for the two
respective considered links. The SC\ expressions in \eqref{Csdl} and %
\eqref{Csul} defines the worst-case SC\ with respect to all malicious
eavesdroppers and users in the network.

\subsection{Radar Sensing Model}

In addition to ensuring a secure transmission with respect to the set of
malicious targets, the FD\ BS desires to perform monostatic radar sensing to
detect the $L\ $malicious targets. Herein, the objective is to establish a
robust beam and RIS\ reflection designs in order to guarantee an
illumination of the $L$ targets by the transmitted signals, composed of the
legitimate data one and the AN. By leveraging the reflected echo signals
back to the BS, after hitting the targets, the BS can solve the target
detection problem, which depends essentially on the received relative echo
signal power from each target. To this end, for assessing the angular
directivity of the reflected sensing beams by $R$, the RIS beampattern
is a viable metric to quantify the level of signal energy beamformed by the
RIS\ in various angular directions, i.e.,
\begin{align}
P_{B}^{(l)}\left( \varphi _{0},\theta _{0}\right) & =\mathbb{E}\left[
\left\vert \mathbf{h}_{B}\left( \varphi _{0},\theta _{0}\right) \mathbf{s}%
\right\vert ^{2}\right] ,  \notag \\
& =\mathbf{h}_{B}\left( \varphi _{0},\theta _{0}\right) \underset{\triangleq
\mathbf{C}_{s}}{\underbrace{\mathbb{E}\left[ \mathbf{ss}^{H}\right] }}%
\mathbf{h}_{B}^{H}\left( \varphi _{0},\theta _{0}\right) ,l=1,\ldots ,L
\label{beampattern}
\end{align}%
Herein, we define
\begin{equation}
\mathbf{h}_{B}\left( \varphi _{0},\theta _{0}\right) =\mathbf{h}_{R}\left(
\varphi _{0},\theta _{0}\right) \mathbf{\Phi} \mathbf{H}_{BR}  \label{hb}
\end{equation}%
with
\begin{equation}
\mathbf{h}_{R}\left( \varphi _{0},\theta _{0}\right) =\mathbf{a}_{R}\left(
\theta _{0},\varphi _{0}\right) .  \label{hr}
\end{equation}%
It can be inferred from (\ref{Ar}), (\ref{A}), (\ref{hb}), and (\ref{hr})
that $\varphi _{0}$ and $\theta _{0}$ represents the azimuth and elevation
look directions of the RIS, where $P_{B}^{(l)}\left( \varphi _{0},\theta
_{0}\right) $ is the amount of beam-steered signal power in the direction
defined by $(\varphi _{0},\theta _{0})$. By utilizing the definition of $%
\mathbf{s}$ in \eqref{sig}, its covariance matrix can be expressed as
\begin{equation}
\mathbf{C}_{s}=\mathbf{C}_{z}+\sum\limits_{j=1}^{J}\mathbf{W}_{j}  \label{Cs}
\end{equation}%
which yields
\begin{equation}
P_{B}^{(l)}\left( \varphi _{0},\theta _{0}\right) =\mathrm{Tr}\left[ \mathbf{%
H}_{B}\left( \varphi _{0},\theta _{0}\right) \mathbf{C}_{s}\right]
,l=1,\ldots ,L,  \label{beampattern2}
\end{equation}%
with $\mathbf{H}_{B}\left( \varphi _{0},\theta _{0}\right) \triangleq
\mathbf{h}_{B}^{H}\left( \varphi _{0},\theta _{0}\right) \mathbf{h}%
_{B}\left( \varphi _{0},\theta _{0}\right) .$ The metric in (\ref%
{beampattern2}) is useful in beamscan applications, where $B$ focuses on
beamsteering a signal beam, with the help of RIS-aided reflection, in an intended direction of interest, defined by
the pair $\left( \varphi _{0},\theta _{0}\right) $ to maximize the target detection performance, given the target is located in the angular direction $%
\left( \varphi _{0},\theta _{0}\right) $ \cite{bazzi2}. Note that unlike the transmit beampattern metric defined for RIS-less radar and ISAC systems, the direction $\left( \varphi _{0},\theta _{0}\right)$ is with respect to $R$'s broadside direction. Herein, by exploiting the knowledge of the RIS location, the optimal configuration of its phase shifts can control the directivity of the beam in the direction(s) of interest, as seen from the perspective of $R$.

\begin{remark}
In the sequel, we consider a two-dimensional beamforming case (the same
elevation angle for all nodes), i.e., $P_{B}^{(l)}\left( \varphi _{0},\theta
_{0}\right) =P_{B}^{(l)}\left( \varphi _{0}\right) $, and we aim at
maximizing $P_{B}^{(l)}$ in the azimuth angles of interest (targets'
locations).
\end{remark}

\begin{remark}
From (\ref{snrevedl}), (\ref{snreveul}), (\ref{ccdleve}), (\ref{cculeve}),
and (\ref{beampattern2}), it can be observed that the AN signal $\mathbf{z}$
not only reduces the channel capacities of the eavesdropping links, but also
increases the signal energy level reaching the sensed targets.
\end{remark}

Another metric for evaluating the efficacy of the sensing scheme is the
total received sensing signal power at the targets planes, defined as%
\begin{equation}
P_{s}^{(l)}=\mathrm{Tr}\left[ \mathbf{H}_{BE_{l}}\mathbf{C}_{s}\right] .
\label{senspw}
\end{equation}%
Based on (\ref{rxsigbs}), we also define the received echo SINR at the BS
for detecting the $l$th target as\
\begin{equation}
\gamma _{\mathrm{echo}}^{(l)}=\frac{\mathbf{f}_{l}^{H}\mathbf{H}_{BE_{l}B}%
\mathbf{C}_{s}\mathbf{H}_{BE_{l}B}^{H}\mathbf{f}_{l}}{\mathbf{f}_{l}^{H}%
\mathbf{R}_l\mathbf{f}_{l}},  \label{snrecho1}
\end{equation}%
which measures the ratio between the power of the echo signal propagated
back to the BS, after hitting $E_{l}$, over the one of other targets and the
communication signals, with
\begin{align}
\mathbf{R}_l& \triangleq \sum\limits_{k=1}^{H}p_{k}\mathbf{h}_{V_{k}B}\mathbf{h%
}_{V_{k}B}^{H}+\sum\limits_{\substack{ m=1 \\ m\neq l}}^{L}\mathbf{H}%
_{BE_{m}B}\mathbf{C}_{s}\mathbf{H}_{BE_{m}B}^{H}  \notag \\
& +\xi _{\mathrm{SI}}\mathbf{H}_{BB}\mathbf{C}_{s}\mathbf{H}_{BB}^{H}+%
\mathbf{R}_{c}+\sigma _{n,B}^{2}\mathbf{I}_{N_{r}}.
\end{align}%
One can note that $\gamma _{\mathrm{echo}}^{(l)}$ is a Rayleigh quotient
expression, which is maximized when
\begin{equation}
\mathbf{f}_{l}^{\text{\textrm{(opt)}}}=\mathrm{maxeigv}\left( \mathbf{R}_l^{-1}%
\mathbf{H}_{BE_{l}B}\mathbf{C}_{s}\mathbf{H}_{BE_{l}B}^{H}\right)
\end{equation}
\begin{remark}
The echo SNR in (\ref{snrecho1}) can be expressed as
\begin{equation}
\gamma _{\mathrm{echo}}^{(l)}=\frac{\rho _{l}\mathbf{f}_{l}^{H}\widetilde{%
\mathbf{h}}_{E_{l}B}\widetilde{\mathbf{h}}_{BE_{l}}\mathbf{C}_{s}\widetilde{%
\mathbf{h}}_{BE_{l}}^{H}\widetilde{\mathbf{h}}_{E_{l}B}^{H}\mathbf{f}_{l}}{%
\mathbf{f}_{l}^{H}\mathbf{R}_l\mathbf{f}_{l}},  \label{snrecho2}
\end{equation}%
where one can observe from the definition of $\widetilde{\mathbf{h}}_{BE_{l}}
$ between (\ref{snrbs}) and (\ref{snreveul}), in terms of $\widetilde{\mathbf{F}}_{BE}
$, that $\widetilde{\mathbf{h}}%
_{BE_{l}}=\mathbf{h}_{BE_{l}}/\sqrt{\mathcal{L}_{BRE_{l}}}$. Thus, by
utilizing the definition $\mathbf{H}_{BE_{l}} = \mathbf{h}_{BE_{l}}^{H}\mathbf{%
h}_{BE_{l}}$ it yields from (\ref{snrecho2}) that%
\begin{equation}
\gamma _{\mathrm{echo}}^{(l)}=\frac{\rho _{l}\mathrm{Tr}\left[ \mathbf{H}%
_{BE_{l}}\mathbf{C}_{s}\right] \mathbf{f}_{l}^{H}\widetilde{\mathbf{h}}%
_{E_{l}B}\widetilde{\mathbf{h}}_{E_{l}B}^{H}\mathbf{f}_{l}}{\mathcal{L}%
_{BRE_{l}}\mathbf{f}_{l}^{H}\mathbf{R}_l\mathbf{f}_{l}},
\end{equation}%
which is proportional to $P_{s}^{(l)}$ in (\ref{senspw}). Therefore, the
latter metric can represent well the sensing performance of an ISAC\
network, where the higher the amount of signal energy reaching a target $%
E_{l}$ $\left( \forall l\in \left\{ 1,\ldots ,L\right\} \right) $, the
higher the reflected echo signal power, resulting in a higher echo SNR.
\end{remark}

\section{Secrecy-Sensing Trade-off Optimization Problem}

In this section, the considered optimization problem optimizing the
secrecy-sensing trade-off of the considered FD-DFRC-RIS network is
formulated, and a solution is presented. In particular, the optimization
problem at hand aims at optimizing the RIS\ phase shifts, BS's DL transmit
beamforming vectors, AN\ covariance matrix, BS's receive combining vectors,
and UL users' transmit power, while ensuring a maximal sensing
performance subject to certain secrecy constraints in both UL and\ DL as
well as a total power constraint.

\subsection{Optimization Problem Formulation and Equivalent Representation}

The considered scheme aims at balancing the sensing-secrecy trade-off in the
considered FD-DFRC-RIS system. From a system design perspective, this
implies the optimal selection of the legitimate signal beamforming matrix $%
\mathbf{W}$, the AN\ covariance matrix $\mathbf{C}_{z}$, the UL users'
receive combining vectors $\left\{ \mathbf{r}_{k}\right\} _{k=1}^{K}$, the
UL users' transmit power vector $\mathbf{p\triangleq }\left[ p_{1},\ldots
,p_{K}\right] $, and the RIS\ phase shifts $\mathbf{\Phi }$, with the
objective of maximizing the secrecy (sensing) performance subject to a
sensing (secrecy) and total power constraint(s). The RIS controller is
connected with the BS\ in order to enable a centralized optimization of the
aforementioned control variables at $B$. In the proposed design, sensing is
prioritized, and the objective function is the received signal power at the
targets' plane, defined by (\ref{senspw}), while ensuring minimal UL\ and
DL\ SC levels. Furthermore, without loss of generality, we assume a perfect
CSI\ available at the DFRC BS, which aligns with the assumption of
stationary nodes in the network and sufficiently longer training pilot
symbols for channel estimation. Thus, the proposed optimization problem is
formulated as 
\begin{subequations}
\label{P1}
\begin{align}
P_{1}& :\max_{\left\{ \mathbf{r}_{k}\right\} _{k=1}^{K},\mathbf{p,}\left\{
\mathbf{W}_{j}\right\} _{j=1}^{J}\mathbf{,C}_{z},\mathbf{\Phi }%
}\min_{l=1,\ldots ,L}P_{s}^{(l)} \\
\text{s.t.}\ (\mathrm{C1})& :C_{s,\mathrm{DL}}\geq C_{\mathrm{DL}}^{(\min )}
\label{C1a} \\
(\mathrm{C2})& :C_{s,\mathrm{UL}}\geq C_{\mathrm{UL}}^{(\min )}  \label{C2a}
\\
(\mathrm{C3})& :\mathrm{Tr}\left[ \mathbf{C}_{s}\right] +\sum%
\limits_{k=1}^{K}p_{k}\leq P_{\max }  \label{C3a} \\
(\mathrm{C4})& :\mathbf{p\succeq 0}  \label{C4a} \\
(\mathrm{C5})& :\mathbf{W}_{j}\mathbf{\succeq 0} \left(\text{ }j=1,\ldots ,J\right), \mathbf{C}_{z}\mathbf{\succeq 0,}
\label{C5a} \\
(\mathrm{C6})& :\mathrm{rank}\left( \mathbf{W}_{j}\right) =1\mathbf{,}\text{
}j=1,\ldots ,J  \label{C6a} \\
(\mathrm{C7})& :\ \left\vert \left[ \mathbf{\Phi }\right] _{n,n}\right\vert
=1  \label{C7a}
\end{align}%
\end{subequations}
%
%
%
where the objective function represents the minimal received signal power
among the $L$ malicious targets (i.e., max-min fairness optimization
problem), while $\mathrm{(C1)}$ and $\mathrm{(C2)}$ are, respectively, the
DL\ and UL\ channels' SC\ constraints, whereby minimal SC\ levels, i.e., $C_{%
\mathrm{DL}}^{(\min )}$ and $C_{\mathrm{UL}}^{(\min )}$, respectively, are
imposed. By properly setting minimal secrecy rate thresholds, such
constraints ensure a minimal information leakage to the malicious
eavesdroppers. Furthermore $\mathrm{(C3)}$ represents the maximal network's
power constraint, with $P_{\max }$ denoting the power budget, while $\mathrm{%
(C4)}$-$\mathrm{(C6)}$ force the non-negativity of the UL\ power levels, the positive semidefinitness property of the matrices $%
\left\{ \mathbf{W}_{j}\right\} _{j=1}^{J}$, $\mathbf{C}_z$, and the unit-rank property of the former set of matrices. Lastly, $\mathrm{(C7)}$ is
linked to the RIS\ reflection coefficients' characteristics, i.e., $%
\left\vert e^{j\theta _{n}}\right\vert =1$ $\forall n$. The aforementioned
problem is non-convex due to $\mathrm{(C6)}$, $\mathrm{(C7)}$, in addition
to $\mathrm{(C1)}$ and $\mathrm{(C2)}$ due to the presence of fractional
expressions in the SC formulas in terms of $\mathbf{W}_{j}$, $\mathbf{C}_{z}$%
, and $p_{k}$, as noticed from the SCs\ and channel capacities definitions
in (\ref{Csdl})-(\ref{cculeve}) and the respective SINRs given in (\ref%
{snruj}), (\ref{snrevedl}), (\ref{snrbs}), and (\ref{snreveul}).

As a first step in reformulating $P_{1}$ into a tractable form, the
following equivalent optimization problem is proposed
\begin{subequations}
\label{P2}
\begin{align}
P_{2}& :\max_{\left\{ \mathbf{r}_{k}\right\} _{k=1}^K,\mathbf{p,}%
\left\{ \mathbf{W}_{j}\right\} _{j=1}^J\mathbf{,C}_{z},\mathbf{\Phi} ,\alpha
}\alpha  \\
\text{s.t.}\ (\mathrm{C1})& :\gamma _{U_{j}}\geq \gamma _{U,\mathrm{DL}}^{%
\text{\textrm{(min)}}},\forall j,  \label{C1b} \\
(\mathrm{C2})& :\gamma _{E_{l},\mathrm{DL}}^{(j)}\leq \gamma _{E,\mathrm{DL}%
}^{\text{\textrm{(max)}}},\forall j,\forall l,  \label{C2b} \\
(\mathrm{C3})& :\gamma _{B,k}\geq \gamma _{B}^{\text{\textrm{(min)}}%
},\forall k,  \label{C3b} \\
(\mathrm{C4})& :\gamma _{E_{l},\mathrm{UL}}^{(k)}\leq \gamma _{E,\mathrm{UL}%
}^{\text{\textrm{(max)}}},\forall k,\forall l,  \label{C4b} \\
(\mathrm{C5})& :P_{s}^{(l)}\geq \alpha ,\text{ }l=1,\ldots ,L  \label{C5b} \\
& \eqref{C3a}-\eqref{C7a}
\end{align}
\end{subequations}
In $P_{2}$, a slack variable $\alpha $ is incorporated to provide an
alternative tractable expression of the challenging objective function of $P_{1}$ in (\ref%
{P1}), expressed as the minimum of received sensing power levels over the $L$ targets.
The introduction of $\alpha $ serves also as a means to ensure a maximal
sensing power for each of the $L$ targets, as represented by (\ref{C5b}).
Furthermore, note that the SC\ constraints in (\ref{C1a})-(\ref{C2a}) are
relaxed now into minimal and maximal SINR levels for the legitimate and
illegitimate channels, respectively, as shown in (\ref{C1b})-(\ref{C4b})
\cite[Eq. (28)]{bazzi}. In the above, $\gamma _{U,\mathrm{DL}}^{\text{%
\textrm{(min)}}}$ and $\gamma _{B}^{\text{\textrm{(min)}}}$ represent the
minimal SINR\ imposed on each DL\ user and the DFRC BS, respectively, while $%
\gamma _{E,\mathrm{DL}}^{\text{\textrm{(max)}}}$ and $\gamma _{E,\mathrm{UL}%
}^{\text{\textrm{(max)}}}$ are the maximal allowable received SINRs\ at each
eavesdropper for decoding DL\ and UL\ signals, respectively. An analogy
between $P_{1}$ and $P_{2}$ can be noted from (\ref{C1b}), (\ref{C2b}),
whereby we can define $C_{U,\mathrm{DL}}^{\text{\textrm{(min)}}}\triangleq
\log _{2}\left( 1+\gamma _{U,\mathrm{DL}}^{\text{\textrm{(min)}}}\right) $
and $C_{E,\mathrm{DL}}^{\text{\textrm{(max)}}}\triangleq \log _{2}\left(
1+\gamma _{E,\mathrm{DL}}^{\text{\textrm{(max)}}}\right) $ as the
corresponding minimal DL legitimate channel capacity and maximal wiretap one,
respectively. One can note that the inequalities of (\ref{C1b}) and (\ref%
{C2b}) give a lower-bound for the legitimate channel capacity $\left( C_{U,\mathrm{%
DL}}^{\text{\textrm{(min)}}}\right) $ and an upper-bound for the wiretap
channel's $\left( C_{E,\mathrm{DL}}^{\text{\textrm{(max)}}}\right) $, which
produces the following
\begin{equation}
\underset{\triangleq C_{s,DL}^{\left( j,l\right) }}{\underbrace{%
C_{U_{j}}-C_{E_{l}\mathrm{,DL}}^{(j)}}}\geq C_{\mathrm{DL}}^{\text{\textrm{%
(min)}}},\forall j,l,  \label{secineq}
\end{equation}%
where $C_{\mathrm{DL}}^{\text{\textrm{(min)}}}\triangleq C_{U,\mathrm{DL}}^{%
\text{\textrm{(min)}}}-C_{E,\mathrm{DL}}^{\text{\textrm{(max)}}}$. Note that
(\ref{secineq}) can be equivalently written as $C_{s,\mathrm{DL}}\geq C_{%
\mathrm{DL}}^{\text{\textrm{(min)}}}$ for positive $C_{s,DL}^{\left(
j,l\right) }$ values, per the worst-case SC\ definition in (\ref{Csdl}). One
can observe that this produces an equivalent representation of the initial
DL\ secrecy constraint in (\ref{C1a}), where the minimal SC\ level can be
tuned by varying either the legitimate links' minimal SINRs, i.e., $\gamma
_{U,\mathrm{DL}}^{\text{\textrm{(min)}}}$, or the eavesdropping links'
maximal ones, i.e., $\gamma _{E,\mathrm{DL}}^{\text{\textrm{(max)}}}$. A\
similar procedure can be performed for linking the UL\ SC constraint in (\ref%
{C2a}) with (\ref{C3b}) and (\ref{C4b}). On the other hand, notice that such
a new secrecy representation, given by (\ref{C1b})-(\ref{C4b}), can help in
transforming $\mathrm{(C1)}$ and $\mathrm{(C2)}$ of $P_{1}$ in (\ref{C1a})-(%
\ref{C2a}) into convex expressions. Observe that the problem in (\ref{P2}) is still
non-convex due to (i)\ the coupling between the RIS\ phase shifts matrix $%
\mathbf{\Phi} $ included implicitly in the channel vectors and coefficients $\left\{
\mathbf{h}_{BU_{j}}\right\} _{j=1}^{J}$, $\left\{ \mathbf{h}%
_{BE_{l}}\right\} _{l=1}^{L}$, $\left\{ \mathbf{h}_{V_{k}B}\right\}
_{k=1}^{K}$, $\left\{ \mathbf{h}_{V_{k}U}\right\} _{k=1}^{K}$, $\left\{
\mathbf{h}_{V_{k}E}\right\} _{k=1}^{K}$ and, consequently, in the SINRs\ in (%
\ref{snruj}), (\ref{snrevedl}), (\ref{snrbs}), and (\ref{snreveul}), as well
as (ii)\ the non-convexity of \eqref{C6a} and \eqref{C7a}. However,
by relaxing the unit-rank constraint \eqref{C6a} and reformulating of
the dependence on $\mathbf{\Phi} $ into semidefinite programming (SDP)\ constraints,
one can solve for a given variable, or jointly for set of non-coupled
variables, while the other coupled variables are fixed. Thus, an AO approach
can be implemented for solving iteratively $P_{2}$ with respect to the
various control variables incorporated.

\subsection{Successive Convex Approximation-based Alternating Optimization
for Secure RIS-aided FD-DFRC\ Design}

\subsubsection{Optimal Receive Combining Vector $\left\{ \mathbf{r}%
_{k}\right\} _{k=1}^{K}$ for Given $\mathbf{p,}\left\{ \mathbf{W}%
_{j}\right\} _{j=1}^{J}\mathbf{,C}_{z},\mathbf{\Phi} ,$ and $\protect\alpha $}

The receive combining vectors for the $K$ UL\ users' signals take part only
in the received UL\ SINR\ at the BS\ expressed in (\ref{snrbs}). The
aforementioned SINR is a Rayleigh quotient expression, for which the optimal
$\mathbf{r}_{k}$ value for given $\mathbf{p,}\left\{ \mathbf{W}_{j}\right\}
_{j=1\ldots ,J}\mathbf{,C}_{z},\mathbf{\Phi} ,$and $\alpha $ values can be expressed
as \cite{bazzi}
\begin{equation}
\mathbf{r}_{k}^{\text{\textrm{(opt)}}}=\mathbf{D}^{-1}_{k}\left( \mathbf{p,C}_{s}\right) \mathbf{h}_{V_{k}B}.  \label{uopt}
\end{equation}
with
\begin{align}
\mathbf{D}_{k}\left( \mathbf{p,C}_{s}\right) & \triangleq \sum\limits
_{\substack{ k^{\prime }=1  \\ k^{\prime }\neq k}}^{H}p_{k^{\prime }}\mathbf{%
h}_{V_{k^{\prime }}B}\mathbf{h}_{V_{k^{\prime }}B}^{H}+\mathbf{H}_{BEB}%
\mathbf{C}_{s}\mathbf{H}_{BEB}^{H}  \notag \\
& +\xi _{\mathrm{SI}}\mathbf{H}_{BB}\mathbf{C}_{s}\mathbf{H}_{BB}^{H}+%
\mathbf{R}_{c}+\sigma _{n,B}^{2}\mathbf{I}_{N_{r}}  \label{Dh}
\end{align}
Consequently, by inserting (\ref{uopt}) into (\ref{snrbs}), the optimal $%
\gamma _{B,k}$ value is obtained as
\begin{equation}
\gamma _{B,k}^{\text{\textrm{(opt)}}}=p_{k}\mathbf{h}_{V_{k}B}^{H}\mathbf{D}%
_{k}^{-1}\left( \mathbf{p,C}_{s}\right) \mathbf{h}_{V_{k}B}.
\label{snrbsopt}
\end{equation}%

\subsubsection{Optimal $\mathbf{p,}\left\{ \mathbf{W}_{j}\right\}
_{j=1\ldots ,J}\mathbf{,C}_{z}$, and $\protect\alpha $ For Given $\left\{
\mathbf{r}_{k}\right\} _{k=1}^{K},\mathbf{\Phi} $}

Once the optimal $\mathbf{r}_{k}$ values are determined, one can plug (\ref%
{snrbsopt}) into (\ref{P2}) and use the definitions of $\gamma _{U_{j}}$, $%
\gamma _{E_{l},\mathrm{DL}}^{(j)}$, $\gamma _{E_{l},\mathrm{UL}}^{(k)}$, and
$P_{s}^{(l)}$ in (\ref{snruj}), (\ref{snrevedl}), (\ref{snreveul}), and (\ref%
{senspw}), respectively to produce the problem $P_{3}$, shown in (\ref{P3})
at the top of the next page. With the relaxation of the unit-rank constraint
on $\mathbf{W}_{j}$, the objective function and the constraints of $P_{3}$
are convex, except for $\mathrm{C3}$. As a remedy to this, the SCA\
technique is utilized to convexify $\mathrm{C3}$. In this approach, a
surrogate convex function of $\mathbf{p,}\left\{ \mathbf{W}_{j}\right\}
_{j=1\ldots ,J}\mathbf{,}$and\textbf{\ }$\mathbf{C}_{z}$ is used to
approximate the non-convex function of the same aforementioned variables in (%
\ref{C3c}). Then, the problem can be solved iteratively by refining at each iteration the constraint function based on the previous iteration's
solution, i.e. $\mathbf{p}^{(n-1)}\mathbf{,}\left\{ \mathbf{W}%
_{j}^{(n-1)}\right\} _{j=1\ldots ,J}$, and $\mathbf{C}_{z}^{(n-1)}$, where $%
n $ is the SCA iteration index \cite{surveyris}.
\begin{figure*}[t]
{\normalsize 
\setcounter{mytempeqncnt}{\value{equation}}
}
\par
\begin{subequations}
\label{P3}
\begin{align}
P_{3}& :\max_{\mathbf{p,}\left\{ \mathbf{W}_{j}\right\} _{j=1}^J%
\mathbf{,C}_{z},\alpha }\alpha \\
\text{s.t.}\ (\mathrm{C1})& :\mathrm{Tr}\left[ \mathbf{H}_{BU_{j}}\mathbf{W}%
_{j}\right] -\gamma _{U,\mathrm{DL}}^{\text{\textrm{(min)}}}\left(
\sum\limits_{\substack{ j'=1  \\ j'\neq j}}^{J}\mathrm{Tr}\left[ \mathbf{H}%
_{BU_{j}}\mathbf{W}_{j'}\right] +\mathrm{Tr}\left[ \mathbf{H}_{BU_{j}}\mathbf{%
C}_{z}\right] +\sum\limits_{k=1}^{K}p_{k}\left\vert
h_{V_{k}U_{j}}\right\vert ^{2}+\sigma _{n,U}^{2}\right) \geq 0,\forall j,
\label{C1c} \\
(\mathrm{C2})& :\mathrm{Tr}\left[ \mathbf{H}_{BE_{l}}\mathbf{W}_{j}\right]
-\gamma _{E,\mathrm{DL}}^{\text{\textrm{(max)}}}\left( \sum\limits
_{\substack{ j'=1  \\ j'\neq j}}^{J}\mathrm{Tr}\left[ \mathbf{H}_{BE_{l}}%
\mathbf{W}_{j'}\right] +\mathrm{Tr}\left[ \mathbf{H}_{BE_{l}}\mathbf{C}_{z}%
\right] +\sum\limits_{k=1}^{K}p_{k}\left\vert h_{V_{k}E_{l}}\right\vert
^{2}+\sigma _{n,E}^{2}\right) \leq 0,\forall j,\forall l,  \label{C2c} \\
(\mathrm{C3})& :p_{k}\mathbf{h}_{V_{k}B}^{H}\mathbf{D}_{k}^{-1}\left(
\mathbf{p},\mathbf{C}_{s}\right) \mathbf{h}_{V_{k}B}\geq \gamma _{B}^{\text{%
\textrm{(min)}}},\forall k,  \label{C3c} \\
(\mathrm{C4})& :p_{k}\left\vert h_{V_{k}E_{l}}\right\vert ^{2}-\gamma _{E,%
\mathrm{UL}}^{\text{\textrm{(max)}}}\left( \sum\limits_{\substack{ k=1  \\ %
k\neq h}}^{K}p_{k}\left\vert h_{V_{k}E_{l}}\right\vert ^{2}+\mathrm{Tr}\left[
\mathbf{H}_{BE_{l}}\mathbf{C}_{s}\right] +\sigma _{n,E}^{2}\right) \leq
0,\forall k,\forall l,  \label{C4c} \\
& \eqref{C3a}-\eqref{C6a} \label{C5c}, 
\end{align}
\end{subequations}
\par
{\normalsize 
\hrulefill 
\vspace*{1pt} }
\end{figure*}

As a first step of the proposed\ SCA technique, we rewrite $\mathrm{C3}$ in (%
\ref{C3c}) in an approximate form by virtue of the Taylor series expansion
around a given point $\mathbf{D}_{k}\left( \mathbf{p,C}_{s}\right) =\mathbf{D%
}_{k}^{(n)}\left( \mathbf{p,C}_{s}\right) $, as
\begin{equation}
p_{k}\mathbf{h}_{V_{k}B}^{H}\mathbf{D}_{k}^{-1}\left( \mathbf{p},\mathbf{C}%
_{s}\right) \mathbf{h}_{V_{k}B}\approx p_{k}\left[
\begin{array}{l}
\mathbf{h}_{V_{k}B}^{H}\left[ \mathbf{D}_{k}^{(n)}\right] ^{-1}\mathbf{h}%
_{V_{k}B} \\
-\mathbf{h}_{V_{k}B}^{H}\left[ \mathbf{D}_{k}^{(n)}\right] ^{-1}\left(
\mathbf{D}_{k}-\mathbf{D}_{k}^{(n)}\right) \\
\times \left[ \mathbf{D}_{k}^{(n)}\right] ^{-1}\mathbf{h}_{V_{k}B}%
\end{array}%
\right]  \label{approxtaylor}
\end{equation}%
where $\mathbf{D}_{k}^{(n)}\left( \mathbf{p,C}_{s}\right) \triangleq \mathbf{%
D}_{k}\left( \mathbf{p}^{(n-1)}\mathbf{,C}_{s}^{(n-1)}\right) $ is the
matrix $\mathbf{D}_{k}$ in (\ref{Dh}) evaluated at the past iteration's
solutions. In (\ref{approxtaylor}) and for the remainder of the paper, the
notation $\mathbf{D}_{k}^{(n)}$ is adopted instead of $\mathbf{D}%
_{k}^{(n)}\left( \mathbf{p,C}_{s}\right) $ for convenience. By arranging the
surrogate function in the right side of (\ref{approxtaylor}) and
substituting it in (\ref{C3c}), one obtains the new reformulated constraint
\begin{equation}
\mathrm{(C3b):}\underset{\triangleq f_{h}\left( \mathbf{p,}\left\{ \mathbf{W}%
_{j}\right\} _{j=1}^J\mathbf{,C}_{z}\right) }{\underbrace{%
\begin{array}{c}
\frac{\gamma _{B}^{\text{\textrm{(min)}}}}{p_{k}}-2\mathbf{h}_{V_{k}B}^{H}%
\left[ \mathbf{D}_{k}^{(n)}\right] ^{-1}\mathbf{h}_{V_{k}B} \\
+\mathbf{h}_{V_{k}B}^{H}\left[ \mathbf{D}_{k}^{(n)}\right] ^{-1}\mathbf{D}%
_{k}\left[ \mathbf{D}_{k}^{(n)}\right] ^{-1}\mathbf{h}_{V_{k}B}%
\end{array}%
}}\geq 0,\forall k,\forall n.  \label{C3d}
\end{equation}

\begin{theorem}
The constraint $\mathrm{(C3b)}$ in (\ref{C3d}) is a convex function of $%
\mathbf{p,}\left\{ \mathbf{W}_{j}\right\} _{j=1}^J$, and $\mathbf{C}%
_{z}$.

\begin{proof}
The convexity of (\ref{C3d}) can be proven in a similar way as performed in
\cite[Eqs. (38)-(41)]{bazzi}.
\end{proof}
\end{theorem}

By substituting (\ref{C3c}) of $P_{3}$ by (\ref{C3d}), an iterative convex
optimization problem is obtained as
\begin{subequations}
\label{P4m}
\begin{align}
P_{4}^{(n)}& :\max_{\mathbf{p,}\left\{ \mathbf{W}_{j}\right\} _{j=1}^J%
\mathbf{,C}_{z},\alpha }\alpha \\
\text{s.t.}\ (\mathrm{C1})& :\frac{\gamma _{B}^{\text{\textrm{(min)}}}}{p_{k}%
}-2\mathbf{h}_{V_{k}B}^{H}\left[ \mathbf{D}_{k}^{(n)}\right] ^{-1}\mathbf{h}%
_{V_{k}B}  \notag \\
& +\mathbf{h}_{V_{k}B}^{H}\left[ \mathbf{D}_{k}^{(n)}\right] ^{-1}\mathbf{D}%
_{k}\left[ \mathbf{D}_{k}^{(n)}\right] ^{-1}\mathbf{h}_{V_{k}B}^{H}\geq
0,\forall k,  \label{C3dd} \\
& \eqref{C1c},\eqref{C2c},\eqref{C4c}, \eqref{C5c}.  \label{C8d}
\end{align}
\end{subequations}
The problem can be iteratively solved by any convex optimization tool (e.g.,
CVX), for a given RIS\ reflection pattern, over a maximal number of
iterations $(\mathcal{N}_{\mathrm{it}})$. Lastly, due to the rank-one
characteristic of $\mathbf{W}_{j}$, it is crucial to express the obtained
solution for each $\mathbf{W}_{j}$, i.e, $\mathbf{W}_{j}^{\mathrm{(opt)}}$,
as rank-one matrix, which can be defined only by an outer product of two
vectors. Among the well-known approaches for transforming an SDP solution's
into a unit-rank matrix, one can approximate $\mathbf{W}_{j}^{\mathrm{(opt)}%
} $ by virtue of its largest eigenvalue and its corresponding eigenvector as
$\mathbf{W}_{j}^{\mathrm{(opt)}}\approx \widetilde{\mathbf{w}}_{j}\widetilde{%
\mathbf{w}}_{j}^{H}$, where $\widetilde{\mathbf{w}}_{j}=\sqrt{\varrho _{\max
}}\mathbf{e}_{j}$ and $\mathbf{e}_{j}\triangleq $\textrm{maxeigv}$\left(
\mathbf{W}_{j}^{\mathrm{(opt)}}\right) \in
\mathbb{C}
^{N_{t}\times 1}$ is the eigenvector of\textbf{\ }$\mathbf{W}_{j}^{\mathrm{%
(opt)}}$ corresponding to its highest eigenvalue $(\varrho _{\max })$.

\subsubsection{Optimal RIS$\ $Phase Shifts $\mathbf{\Phi }$ For Given $%
\mathbf{p,}\left\{ \mathbf{W}_{j}\right\} _{j=1}^{J}\mathbf{,C}_{z},$ and $%
\left\{ \mathbf{r}_{k}\right\} _{k=1}^{K}$}

The RIS\ phase shifts, represented by the diagonal matrix $\mathbf{\Phi }$,
is included in the different links' channel vectors and matrices, i.e., $\mathbf{H}_{XY}=\sqrt{\mathcal{L}_{XRZ}}\mathbf{H}_{RZ}\mathbf{\Phi H}_{XR}$ for $XZ\in \left\{
BU_{j},BE_{l},V_{k}B,V_{k}U_{j},V_{k}E_{l}\right\} $, which consequently
takes part of the legitimate and illegitimate links' SINRs given by (\ref%
{snruj}), (\ref%
{snrevedl}), (\ref{snreveul}), (\ref%
{snrbsopt}), and the
RIS beampattern, defined by (\ref{beampattern2}). The main challenge in such a
representation lies in the non-convex constraint imposed on each RE's
complex-valued reflection coefficient, as given in (\ref{C7a}). Thus, the aforementioned cascaded channel vector alternatively can be formulated as
follows
\begin{equation}
\mathbf{h}_{BZ}=\mathbf{q}\overline{\mathbf{H}}_{BZ}, Z \in \{U_j,E_l\}, \forall
j,\forall l , \label{hxy}
\end{equation}%
where $\overline{\mathbf{H}}_{BZ} \in \mathbb{C}^{N \times N_t}$ is defined as%
\begin{equation}
\overline{\mathbf{H}}_{BZ}=\left[ \mathbf{h}_{B_{1}Z}^{T},\ldots ,\mathbf{h}%
_{B_{N_{t}}Z}^{T}\right]  \in \mathbb{C}^{N \times N_t} ,\label{hbrzbar}
\end{equation}%
with $\mathbf{h}_{B_{m}Z}\triangleq \left[ h_{B_{m}R_{1}}h_{R_{1}Z},\ldots
,h_{B_{m}R_{N}}h_{R_{N}Z}\right] \in
\mathbb{C}
^{1\times N}$ is the cascaded channel vector representation
between the BS's $m$th transmit antenna and node $Z$ through the $N$ REs of $%
R$, where the $n$th element of $\mathbf{h}_{B_{m}Z}$ is the product between
the channel coefficients of the following links: (i) $B$'s $m$th-antenna$-R$'s $n$th RE and $(h_{B_mR_n})$ (ii)\
$R$'s $n$th RE$-Z$ $(h_{R_nZ})$. In a similar manner, we can define
the equivalent representations for $\mathbf{h}_{V_{k}B}$, $h_{V_{k}U_{j}}$,
and $h_{V_{k}E_{l}}$, respectively, as follows%
\begin{equation}
\mathbf{h}_{V_{k}B}=\overline{\mathbf{H}}_{V_{k}B}\mathbf{q}^{T}
\label{hvkbeq}
\end{equation}%
\begin{equation}
h_{V_{k}Z}=\mathbf{q}\overline{\mathbf{h}}_{V_{k}Z},Z\in \left\{
U_{j},E_{l}\right\} ,\forall \left( j,k,l\right)  \label{hvkzeq}
\end{equation}%
with $\overline{\mathbf{H}}_{V_{k}B}\triangleq \left[ \mathbf{h}%
_{V_{k}B_{1}}^{T},\ldots ,\mathbf{h}_{V_{k}B_{N_{r}}}^{T}\right] ^{T}\in
\mathbb{C}
^{N_{r}\times N}$, $\mathbf{h}_{V_{k}B_{m}}\triangleq \left[
h_{V_{k}R_{1}}h_{R_{1}B_{m}},\ldots ,h_{V_{k}R_{N}}h_{R_{N}B_{m}}\right] \in
\mathbb{C}
^{1\times N}$, and $\overline{\mathbf{h}}_{V_{k}Z}\triangleq \left[
h_{V_{k}R_{1}}h_{R_{1}Z},\ldots ,h_{V_{k}R_{N}}h_{R_{N}Z}\right] ^{T}\in
\mathbb{C}
^{N\times 1}$. Similarly, $h_{V_kR_n}$ is the channel coefficient for the link $V_k$-$n$th RE of $R$. Such an alternative representation of channel vectors and
coefficients is paramount as it can enable recasting the problem into an
equivalent SDP\ one. Thus, by (i) inserting the representations given by (%
\ref{hxy})-(\ref{hvkzeq}) into the constraints of (\ref{P2}) and (ii)
expressing\ the products involving $\mathbf{q}$ using positive semidefinite matrices
in terms of traces, it yields $P_{5}$ given by (\ref{P5}) at the top of the
page, with $\mathbf{Q}\triangleq \mathbf{qq}^{H}$, $\mathbf{G}%
_{BZ}^{(j)}\triangleq \overline{\mathbf{H}}_{BZ}\mathbf{W}_{j}\overline{%
\mathbf{H}}_{BZ}^{H}$, $\mathbf{R}_{BZ}\triangleq \overline{\mathbf{H}}_{BZ}%
\mathbf{C}_{z}\overline{\mathbf{H}}_{BZ}^{H}$, $\overline{\mathbf{H}}%
_{V_{k}Z}\triangleq \overline{\mathbf{h}}_{V_{k}Z}\overline{\mathbf{h}}%
_{V_{k}Z}^{H}$,\textbf{\ }$\mathbf{S}_{BE_{l}}\triangleq \overline{\mathbf{H}%
}_{BE_{l}}\mathbf{C}_{s}\overline{\mathbf{H}}_{BE_{l}}^{H}$,
\begin{figure*}[t]
{\normalsize 
\setcounter{mytempeqncnt}{\value{equation}}
}
\par
\begin{subequations}
\label{P5}
\begin{align}
P_{5}& :\max_{\mathbf{Q},\alpha }\alpha \\
\text{s.t.}\ (\mathrm{C1})& :\mathrm{Tr}\left[ \mathbf{QG}_{BU_{j}}^{(j)}%
\right] -\gamma _{U,\mathrm{DL}}^{\text{\textrm{(min)}}}\left( \sum\limits
_{\substack{ j'=1  \\ j'\neq j}}^{J}\mathrm{Tr}\left[ \mathbf{QG}%
_{BU_{j}}^{(j')}\right] +\mathrm{Tr}\left[ \mathbf{QR}_{BU_{j}}\right]
+\sum\limits_{k=1}^{K}p_{k}\mathrm{Tr}\left[ \mathbf{Q}\overline{\mathbf{H}}%
_{V_{k}U_{j}}\right] +\sigma _{n,U}^{2}\right) \geq 0,\forall j,  \label{C1e}
\\
(\mathrm{C2})& :\mathrm{Tr}\left[ \mathbf{QG}_{BE_{l}}^{(j)}\right] -\gamma
_{E,\mathrm{DL}}^{\text{\textrm{(max)}}}\left( \sum\limits_{\substack{ j'=1
\\ j'\neq j}}^{J}\mathrm{Tr}\left[ \mathbf{QG}_{BE_{l}}^{(j')}\right] +\mathrm{%
Tr}\left[ \mathbf{QR}_{BE_{l}}\right] +\sum\limits_{k=1}^{K}p_{k}\mathrm{Tr}%
\left[ \mathbf{Q}\overline{\mathbf{H}}_{V_{k}E_{l}}\right] +\sigma
_{n,E}^{2}\right) \leq 0,\forall j,\forall l,  \label{C2e} \\
(\mathrm{C3})& :p_{k}\mathrm{Tr}\left[ \mathbf{Q}^{H}\overline{\mathbf{h}}%
_{V_{k}B}^{H}\mathbf{E}_{k}^{-1}\left( \mathbf{Q}\right) \overline{\mathbf{h}%
}_{V_{k}B}\right] \geq \gamma _{B}^{\text{\textrm{(min)}}},\forall k,
\label{C3e} \\
(\mathrm{C4})& :p_{k}\mathrm{Tr}\left[ \mathbf{Q}\overline{\mathbf{H}}%
_{V_{k}E_{l}}\right] -\gamma _{E,\mathrm{UL}}^{\text{\textrm{(max)}}%
}\left( \sum\limits_{\substack{ k'=1  \\ k'\neq k}}^{K}\mathrm{Tr}\left[
\mathbf{Q}\overline{\mathbf{H}}_{V_{k'}E_{l}}\right] +\mathrm{Tr}\left[
\mathbf{QS}_{BE_{l}}\right] +\sigma _{n,E}^{2}\right) \leq ,\forall
k,\forall l,  \label{C4e} \\
(\mathrm{C5})& :\mathrm{Tr}\left[ \mathbf{QG}_{BE_{l}}\right] \geq \alpha ,%
\text{ }l=1,\ldots ,L  \label{C5e} \\
(\mathrm{C6})& :\ \mathbf{Q}_{n,n}=1  \label{C6e} \\
(\mathrm{C7})& :\ \mathrm{rank}(\mathbf{Q)}=1  \label{C7e}
\end{align}
\end{subequations}
\par
{\normalsize 
\hrulefill 
\vspace*{1pt} }
\end{figure*}
\begin{align}
\mathbf{E}_{k}\left( \mathbf{Q}\right) & =\sum\limits_{\substack{ k'=1  \\ %
k'\neq k}}^{K}p_{k'}\overline{\mathbf{h}}_{V_{k'}RB}\mathbf{Q}^{H}\overline{%
\mathbf{h}}_{V_{k'}RB}^{H}  \notag \\
& +\sum\limits_{l=1}^{L}\mathrm{Tr}\left[ \mathbf{Q}\overline{\mathbf{H}}%
_{BE_{l}}\mathbf{C}_{s}\overline{\mathbf{H}}_{BE_{l}}^{H}\right] \overline{%
\mathbf{H}}_{E_{l}B}\mathbf{Q}^{H}\overline{\mathbf{H}}_{E_{l}B}^{H}  \notag
\\
& +\xi _{\mathrm{SI}}\mathbf{H}_{BB}\mathbf{C}_{s}\mathbf{H}_{BB}^{H}+%
\mathbf{R}_{c}+\sigma _{n,B}^{2}\mathbf{I}_{N_{r}}  \label{EQ}
\end{align}%
and $\overline{\mathbf{H}}_{E_{l}B} \in \mathbb{C}^{N_r \times N}$ is defined similarly to $\overline{%
\mathbf{H}}_{BE_{l}}$ by swapping the transceivers order and substituting $%
N_{t}$ by $N_{r}$. $\mathbf{H}_{BB}$ can also be defined in terms of $\mathbf{Q}$ as $\mathbf{H}_{BB}=\mathbf{H}_{BB}^{\mathrm{(DL)}}+\sqrt{\mathcal{L}_{BRB}}%
\mathbf{H}_{RB} \mathrm{diag}\left(\sqrt{\beta_{\mathbf{Q}}} \mathrm{maxeigv}\left(\mathbf{Q}\right)\right) \mathbf{H}_{BR}$ where $\beta_{\mathbf{Q}}$ is the largest eigenvalue of $\mathbf{Q}$. Note that $\mathbf{Q\in
\mathbb{C}
}^{N\times N}$ is a positive semidefinite matrix of unit rank, representing
the RIS\ complex-valued reflection coefficients. Such a representation\ can
help in alleviating the non-convex modulus constraint of complex RIS
reflection coefficients, i.e., $\left\vert \phi _{n}\right\vert =1$, initially represented by \eqref{C7a}. The latter constraint is transformed into a linear one, given by (\ref{C6e}). However, $%
P_{5}$ still possesses two non-convex constraints, namely the UL SINR's
in (\ref{C3e}) and the unit-rank constraint in (\ref{C7e}), imposed on $%
\mathbf{Q}$ by its nature. It is noted that the nonconvexity of $\mathrm{C3}$
in (\ref{C3e}) stems from the matrix $\mathbf{E}_{h}\left( \mathbf{Q}\right)
$ in (\ref{EQ}). Thus, as a way to convexity this constraint, $P_{5}$ will
be solved iteratively whereby at each iteration of the inner layer, i.e., $%
\mathbf{\Phi} $'s optimization subproblem, (\ref{C3e}) will be substituted by the
following convex constraint
\begin{equation}
(\mathrm{C3}):p_{k}\mathrm{Tr}\left[ \mathbf{Q}^{H}\overline{\mathbf{h}}%
_{V_{k}RB}^{H}\mathbf{E}_{k}^{-1}\left( \mathbf{Q}^{(g-1)}\right) \overline{%
\mathbf{h}}_{V_{k}RB}\right] \geq \gamma _{B}^{\text{\textrm{(min)}}%
},\forall k,  \label{C3ris}
\end{equation}%
where $\mathbf{Q}^{(g-1)}$ is the solution of the previous inner layer's
iteration, and $g$ is the current inner layer's iteration index. Obviously, %
\eqref{C3ris} is a convex SDP expression. Thus, by relaxing the unit-rank
constraint in (\ref{C7e}), $P_{5}$ becomes a convex SDP, which can be solved
by any convex optimization tool (e.g., CVX). The adapted version of $P_{5}$
is given as
\begin{subequations}
\label{P6m}
\begin{align}
P_{6}^{(g)}& :\max_{\mathbf{Q},\alpha }\alpha \\
\text{s.t. }(\mathrm{C1})& :p_{k}\mathrm{Tr}\left[ \mathbf{Q}^{H}\overline{%
\mathbf{h}}_{V_{k}B}^{H}\mathbf{E}_{k}^{-1}\left( \mathbf{Q}^{(g-1)}\right)
\overline{\mathbf{h}}_{V_{k}B}\right] \geq \gamma _{B}^{\text{\textrm{(min)}}%
},\forall k,  \label{C3f} \\
& \eqref{C1e},\eqref{C2e},\eqref{C4e},\eqref{C5e},\eqref{C6e}.
\end{align}
\end{subequations}
Similarly to the matrices $\left\{ \mathbf{W}_{j}\right\} _{j=1}^J$,
the solution of $P_{6}^{(g)}$ $(\mathbf{Q}^{\mathrm{(opt)}})$ needs to be of
unit rank. Such a property can be fulfilled through a Gaussian randomization
method, yielding $\mathbf{q}^{\mathrm{(opt)}}=\mathbf{U\Sigma }^{1/2}\exp
\left( i\mathbf{d}\right) $, where $\mathbf{U}$ is the matrix formed by the
eigenvectors of $\mathbf{Q}^{\mathrm{(opt)}}$, $\mathbf{\Sigma }$ is a
diagonal matrix for which the entries are the eigenvalues of $\mathbf{Q}^{%
\mathrm{(opt)}}$, and $\mathbf{d}$ is a stochastic vector with independent
and uniformly distributed elements in $\left[ -\pi ,\pi \right] $. In
Algorithm \ref{algg}, we provide a pseudocode showing the followed algorithmic approach to solve the overall optimization problem, formed by
the solution in (\ref{uopt}) and the elementary subproblems in (\ref{P4m})
and (\ref{P6m}).

\begin{algorithm}[h]
 \SetAlgoLined
\KwData{$\mathbf{h}_{BR}$, $\mathbf{h}_{RB}$, $\left\{\mathbf{h}_{RU_j} \right\}_{j=1}^J$, $\left\{\mathbf{h}_{RE_l} \right\}_{l=1}^L$, $\left\{\mathbf{h}_{V_kR} \right\}_{k=1}^K$, $\boldsymbol{\rho}$, $\xi_{\mathrm{SI}}$, $\mathbf{H}_{BB}$, $P_{\max}$, $\sigma_{n,U}^2$, $\sigma_{n,E}^2$, $\sigma_{n,B}^2$   $\mathcal{M}_{\mathrm{it}}$, $\mathcal{N}_{\mathrm{it}}$, $\mathcal{G}_{\mathrm{it}}$ }
\KwResult{$\mathbf{p}^{\mathrm{(opt)}}$, $\left\{ \mathbf{W}_j^{\mathrm{(opt)}} \right\}_{j=1}^J$, $\mathbf{C}^{\mathrm{(opt)}}_z$, $\left\{ \mathbf{r}_k^{\mathrm{(opt)}} \right\}_{k=1}^K$, $\mathbf{q}^{\mathrm{(opt)}}$}
 \Begin{
 \string\\ \textcolor{blue}{Initialization} \\
 $m \gets 0$  ,  $\mathbf{C}^{(0)}_z \gets \mathbf{0}_{N_t \times N_t}$ \\
 $\mathbf{H}_{BU} \gets \left[ \mathbf{h}_{BU_1}^T ,\ldots, \mathbf{h}_{BU_J}^T  \right]^T$ \\
 $\mathbf{W}_{\mathrm{ZF}} \gets \mathbf{H}_{BU}^H \left( \mathbf{H}_{BU} \mathbf{H}_{BU}^H \right)^{-1}$ \\
 $\left\{\mathbf{W}_j^{(0)}\right\}_{j=1}^J\gets \left\{\left[ \mathbf{W}_{\mathrm{ZF}} \right]_{:,j} \left[ \mathbf{W}_{\mathrm{ZF}} \right]_{:,j}^H\right\}_{j=1}^J$ \\
 $\mathbf{p}^{(0)} \gets \mathbf{1}_{1 \times K} $ , $ \mathbf{q}^{(0)} \gets -\angle(\mathbf{a}_{R} \left( \theta_{RB},\varphi_{RB}\right))$
  \\

 Compute $\left\{\mathbf{r}^{(\mathrm{opt})}_k \right\}_{k=1,\ldots,K} $ using \eqref{uopt} 

\For{$m\gets1$ \KwTo $\mathcal{M}_{\mathrm{it}}$}
    {
\string\\ \textcolor{blue}{Alternating Optimization loop}

\For{$n\gets1$ \KwTo $\mathcal{N}_{\mathrm{it}}$}
    {
    \string\\ \textcolor{blue}{Subproblem 1}
    \\

    Solve $P_4^{(n)}$ in \eqref{P4m} to obtain $\left\{\mathbf{W}_j^{(m,n)}\right\}_{j=1}^J$, $\mathbf{C}^{(m,n)}_z$, $\mathbf{p}^{(m,n)}$ \\

    $\mathbf{C}_s^{(n)} \gets \sum_{j=1}^{J} \mathbf{W}^{(m,n)}_j + \mathbf{C}_z^{(m,k)}$

    }

$\left\{\mathbf{W}_j^{(m)}\right\}_{j=1}^J \gets \left\{\mathbf{W}_j^{(m,\mathcal{N}_{\mathrm{it}})}\right\}_{j=1}^J$ \\
$\mathbf{C}^{(m)}_z \gets \mathbf{C}^{(m,\mathcal{N}_{\mathrm{it}})}_z$ ,
$\mathbf{p}^{(m)} \gets \mathbf{p}^{(m,\mathcal{N}_{\mathrm{it}})}$ \\
\For{$j \gets 1$ \KwTo $J$}
{
Compute $\varrho_{\max}$ and $\mathbf{a}^{(j)}_{\max}$ \string\\ \textcolor{blue}{maxeigv of $\mathbf{W}_j$} \\
$\mathbf{W}_j \gets \varrho_{\max} \mathbf{a}^{(j)}_{\max} \left(\mathbf{a}^{(j)}_{\max}\right)^H $
}

\For{$g \gets 1$ \KwTo $\mathcal{G}_{\mathrm{it}}$}
    {
    \string\\ \textcolor{blue}{Subproblem 2} \\

    Solve $P_6^{(g)}$ in \eqref{P6m} to obtain $\mathbf{Q}^{(m,g)}$ \\
    Compute $\mathbf{U}$ and $\mathbf{\Sigma}$ for $\mathbf{Q}^{(m,g)} $ 
    $\mathbf{d} \gets \left[d_1,\ldots,d_N\right]^T$ with $d_n \sim \mathcal{U}[-\pi,\pi]$ \\
    $\mathbf{q}_s^{(g,m)} \gets  \mathbf{U} \mathbf{\Sigma}^{1/2} e^{i\mathbf{d}} $
    }

Recompute the channels with $\mathbf{q}_s^{(\mathcal{G}_{\mathrm{it}},m)}$, \eqref{hbrz}, \eqref{hvkr}, \eqref{channelsi} \\

    }
$\left\{\mathbf{W}^{(\mathrm{opt})}\right\}_{j=1}^J \gets \left\{\mathbf{W}^{(\mathcal{M}_{\mathrm{it}})}\right\}_{j=1}^J$ ,
$\mathbf{C}^{(\mathrm{opt})}_z \gets \mathbf{C}^{(\mathcal{M}_{\mathrm{it}})}_z$ ,
$\mathbf{q}_s^{(\mathrm{opt})} \gets \mathbf{q}_s^{(\mathcal{G}_{\mathrm{it}},\mathcal{M}_{\mathrm{it}})}$
}
\caption{Secure FD RIS-aided ISAC design.}
\label{algg}
\end{algorithm}

\section{Numerical Evaluation}

In this section, representative numerical results for the secrecy and
sensing performance of the proposed scheme are presented to showcase the
main system parameters' effect on the proposed scheme's performance. Unless
otherwise stated, Table \ref{sysparam} provides the default values
considered for the various system parameters in the considered network and
optimization framework. Furthermore, it is assumed that all nodes are
positioned in the same $xy$ plane at $z=0$ and the DFRC FD BS is positioned
at the origin of the considered reference Cartesian frame, i.e., $x_B=y_B=0$%
. Each category of nodes, namely UL users, DL users, and eavesdroppers are
distributed in an angular equidistant fashion over their respective angular
intervals given in Table \ref{sysparam}. Thus, the inter-node and RIS-nodes
distances can be evaluated using their computed Cartesian coordinates based
on the angle and distance with respect to the BS. Additionally, the noise
power was evaluated using the thermal noise model, i.e. $%
\sigma_{n,Z}^2=N_F k_B T B $, ($Z \in \{B,U,E\}$), with $N_F$ is the
receiver's noise figure, $k_B$ is the Boltzmann constant, $T$ is the
receiver's temperature, and $B$ is the receiver's bandwidth. The value of $\lambda$ corresponds to a 3.5-GHz carrier frequency used in the frequency range 1 (FR1) of the 5G \cite{5G1}. The absence of a direct link can take place in the presence of multiple layers of shadowing objects. For instance, a concrete wall can provide a loss exceeding $30$ dB, whereas a receiver or target obstructed by three walls can suffer from a minimal penetration loss of $90$ dB which, in addition to the distance-dependent attenuation (computed with the distances in Table \ref{sysparam}), can yield a total attenuation of $-123$ dB, which is below the noise floor \cite{ETSI,5G2}.

\begin{table}[t]
\caption{System parameters' values}\centering%
\begin{tabular}{|c|c|}
\hline\hline
\textbf{Parameter} & \textbf{Value/Range} \\ \hline\hline
$\lambda$ & $8.5$ cm \\ \hline
$G_{R,U}$, $G_{R,E}$ & $12$ dBi \\ \hline
$G_{T,B}$, $G_{R,B}$ & $25$ dBi \\ \hline
$G_{T,V}$ & $17$ dBi \\ \hline
$d_{BR}$ & $22$ m \\ \hline
$d_{BU_{j}}(\forall j)$ & $30$ m \\ \hline
$d_{BE_{l}},d_{BV_{k}}(\forall l,k)$ & $20$ m \\ \hline
$\Delta _{r},\Delta _{a}$ & $\lambda /2$ \\ \hline
$\varphi _{BU}$ & $[-15,5]$ deg \\ \hline
$\varphi _{VB}$ & $[20,30]$ deg \\ \hline
$\varphi _{BR}$ & $-40$ deg \\ \hline
$\varphi _{BE}$ & $[-35,-15]$ deg \\ \hline
$\theta _{XZ}$ $\left(\forall X,Z\right)$ & $0$ \\ \hline
$N$ & $25$ ($5\times 5$ REs) \\ \hline
$N_{F}$ & $5$ dB \\ \hline
$P_{\max }$ & $25$ dB \\ \hline
$N_{t}$, $N_{r}$ & $8$ \\ \hline
$B$ & $50$ MHz \\ \hline
$T$ & $298$ Kelvins \\ \hline
$J,L$ & $2$ \\ \hline
$K$ & $3$ \\ \hline
$\gamma _{E,\mathrm{DL}}^{(\max )}$, $\gamma _{E,\mathrm{DL}}^{(\max )}$ & $%
5 $ dB \\ \hline
$\left( \gamma _{B}^{(\min )},\gamma _{U,\mathrm{DL}}^{(\min )}\right) $ & $%
(5,10)$ dB \\ \hline
$\kappa _{XZ}$ $\left(\forall XZ \in \left\{BR, RU_j, V_kR \right\}\right)$ & $20$ dB \\ \hline
$\kappa _{RE_l}$ $\left(\forall l \right)$ & $\infty$ (Pure LoS) \\ \hline
\end{tabular}
\label{sysparam}
\end{table}


\begin{figure}[tbp]
\begin{center}
\includegraphics[scale=.48]{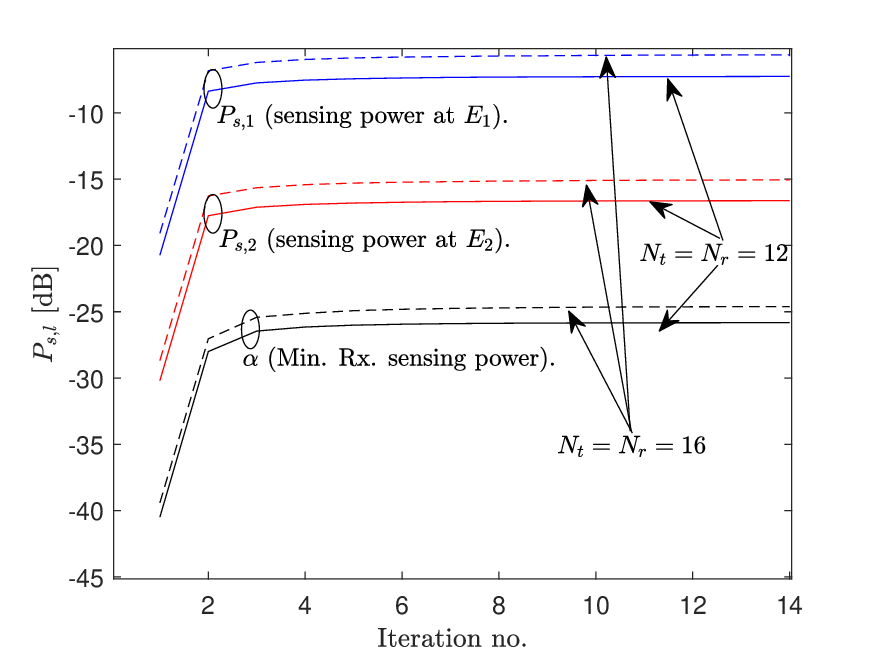}
\end{center}
\par
\vspace*{-.2cm}
\caption{{Evolution of $P_s^{(l)}$ at the eavesdroppers'
directions and the objective function of } $P_{6}^{(q)}$ ($\protect\alpha $)
vs. the iteration index.}
\label{fig1}
\end{figure}

In Fig. \ref{fig1}, the convergence of the received sensing power at the
eavesdroppers' plane is plotted vs. the iteration number of the proposed
AO-based algorithm in Algorithm \ref{algg}. We also show the evolution of
the objective function of \eqref{P6m} ($\alpha $), representing a lower bound for the
received sensing power at the targets, as manifested by \eqref{C5e}. Note
that the proposed approach converges to a certain value after a few number
of iterations, validating the proposed approach for jointly optimizing the
considered control variables. In addition, the power levels at the
eavesdropper's plane exceed their respective lowerbound $\alpha $ per %
\eqref{C5e}. Furthermore, the increase in the number of transmit and receive
antennas yields an increase in the received sensing signal power due to the
increase of the BS's transmit beamforming gain towards the RIS. Therefore, a
higher signal power is reflected to each of the two targets. In particular,
a $2$-dB increase of the received sensing power is noticed at each target
by increasing $N_{t}$ from $12$ to $16$.


\begin{figure}[tbp]
\begin{center}
\includegraphics[scale=.48]{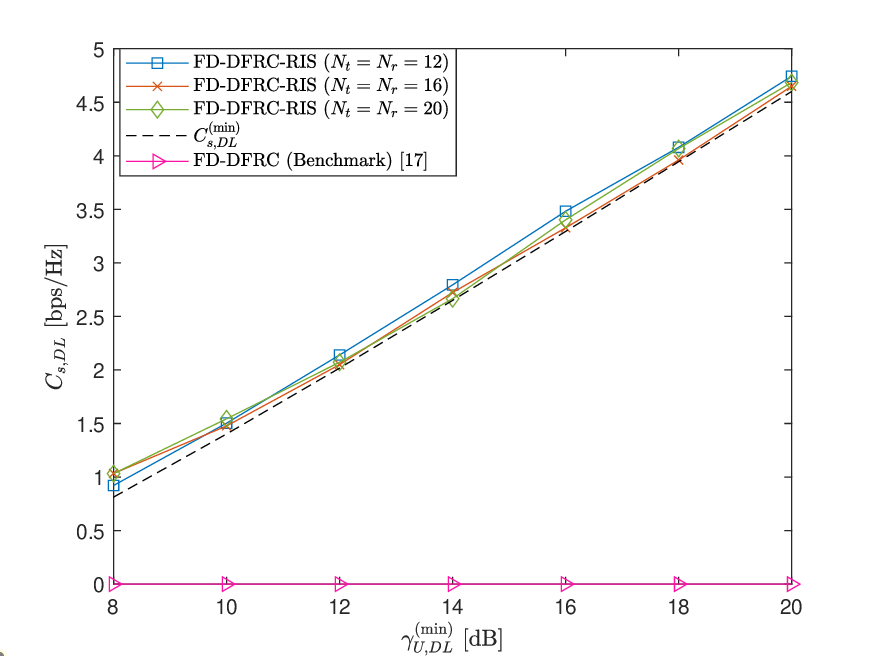}
\end{center}
\par
\vspace*{-.2cm}
\caption{DL SC of the considered scheme vs. $\protect\gamma _{U,\mathrm{DL}%
}^{(\min )}$ evaluated for different $N_{t},N_{r}$ values.}
\label{fig2}
\end{figure}
In Fig. \ref{fig2}, the DL SC of the proposed RIS-aided scheme is shown for
three different $N_{t},N_{r}$ values, namely $N_{t}=N_{r}=12$, $%
N_{t}=N_{r}=16$, and $N_{t}=N_{r}=20$. We set $\kappa _{XZ}=15$ dB $
\left(\forall XZ \in \left\{BR, RU_j, V_kR \right\}\right) $. One can observe that the achievable DL SC of the
proposed scheme yields an independent value of the number of antennas, where
it equals or slightly surpasses the minimal SC requirement defined by: $C_{%
\mathrm{DL}}^{\text{\textrm{(min)}}}=\log _{2}\left( 1+\gamma _{U,DL}^{(\min
)}\right) -\log _{2}\left( 1+\gamma _{E,DL}^{(\max )}\right) $. This is due
to the fact that the DL secrecy performance of the system is set as a
constraint, as given by (\ref{C1e}) and (\ref{C2e}) of the
manuscript. Thus, the proposed framework focuses on fulfilling the
aforementioned DL\ SC\ requirement while maximizing the network's sensing
performance. In particular, a level of $4.6$ bps/Hz of worst-case DL SC can be
achieved, specified by the following DL SINR requirements: $\gamma
_{U,DL}^{(\min )}=20$ dB and $\gamma _{E,DL}^{(\max )}=5$ dB. Of note, the
benchmark scheme in \cite{bazzi} yields a zero SC independently of the SINR
requirements. This is due to the fact that the FD ISAC design \cite{bazzi}
is implemented in the absence of the RIS. Thus, in the absence of direct
communication links between the BS and the various nodes, the SINRs at the
legitimate receivers and eavesdroppers are set to $0$, yielding a zero SC.\
Consequently, the \textit{proposed RIS-aided FD-DFRC design in suitable in
scenarios of a direct link absence between the DFRC\ BS\ and the various
targets and communication nodes in the network. \ }

\begin{figure}[tbp]
\centering
\includegraphics[scale=.48]{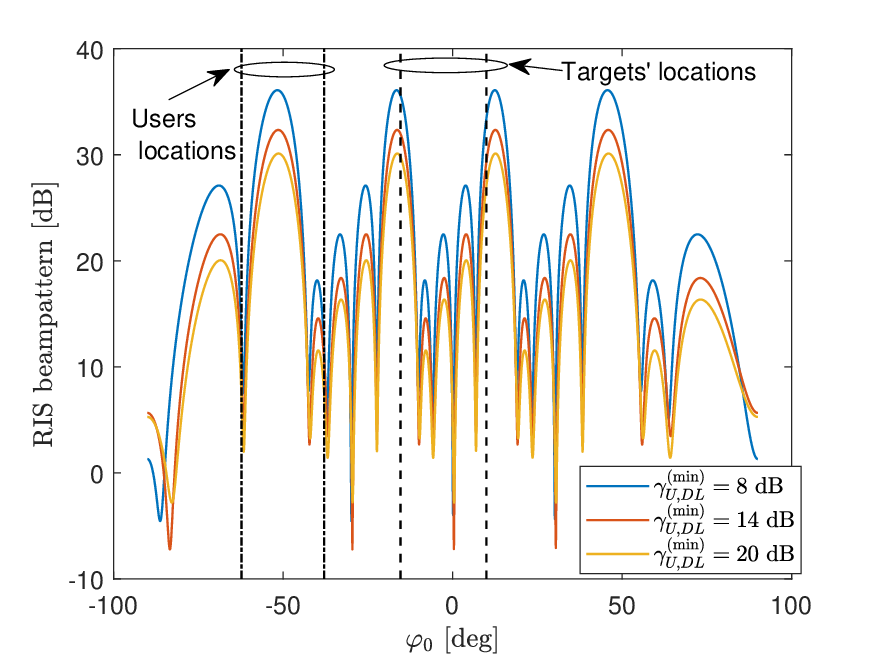} \vspace*{%
-.2cm}
\caption{RIS beampattern vs. the RIS look-direction $(%
\protect\varphi _{0})$ for three different DL SINR requirements.}
\label{fig3}
\end{figure}


In Fig. \ref{fig3}, the RIS beampattern, expressed by (\ref%
{beampattern2}), is plotted vs. the RIS azimuth look-direction ($\varphi
_{0} $) for $N_{t}=N_{r}=12$ and three different $\gamma _{U,DL}^{(\min )}$
values, namely $\gamma _{U,DL}^{(\min )}=8$, $12$, and $20$ dB. Similarly to
Fig. \ref{fig2}, we set $\kappa _{XZ}=15$ dB $
\left(\forall XZ \in \left\{BR, RU_j, V_kR \right\}\right) $. Several observations can be noticed. For
instance, one can observe that the proposed scheme focuses its reflected
beams towards the malicious targets, whereby a higher signal power is
dedicated to $E_{2}$, located farther to the RIS than $E_{1}$, which is
significantly higher than the one directed towards the users. In fact, the
AN\ power is minimized in the direction of the legitimate users and
maximized in the directions of Eves, yielding a higher total sensing power
in the direction of the latter nodes. Furthermore, the increase of the
legitimate DL\ SINR\ requirement reduces the reflected signal power in the
direction of the malicious targets'. It can be inferred from (\ref{snruj})
that the increase of $\gamma _{U_{j}}$ can be controlled by either
increasing the magnitude of $\mathbf{W}_{j}$ or decreasing the one of $%
\mathbf{C}_{z}$. The increase in the former yields a higher SINR at the
illegitimate receivers, potentially violating (\ref{C2c}). To remedy to
this, a simultaneous increase of $\mathbf{C}_{z}$'s magnitude along with $%
\mathbf{W}_{j}$'s can fulfill the eavesdropping SINR constraints in (\ref%
{C2c}), which might violate the total power constraint in (\ref{C3a}). Thus,
the proposed\ AO-based scheme favoritizes reducing the legitimate signal
power with a further reduction on the AN\ power to guarantee a higher DL\
SINR requirement. Thus, according to (\ref{Cs}) and (\ref{beampattern2}),
this reduces the total energy beamformed towards the target. Therefore, the
realization of a more stringent secure communication trades off with the
amount of legitimate signal leakage and AN\ power invested on sensing.
\begin{figure}[tbp]
\begin{center}
\includegraphics[scale=.48]{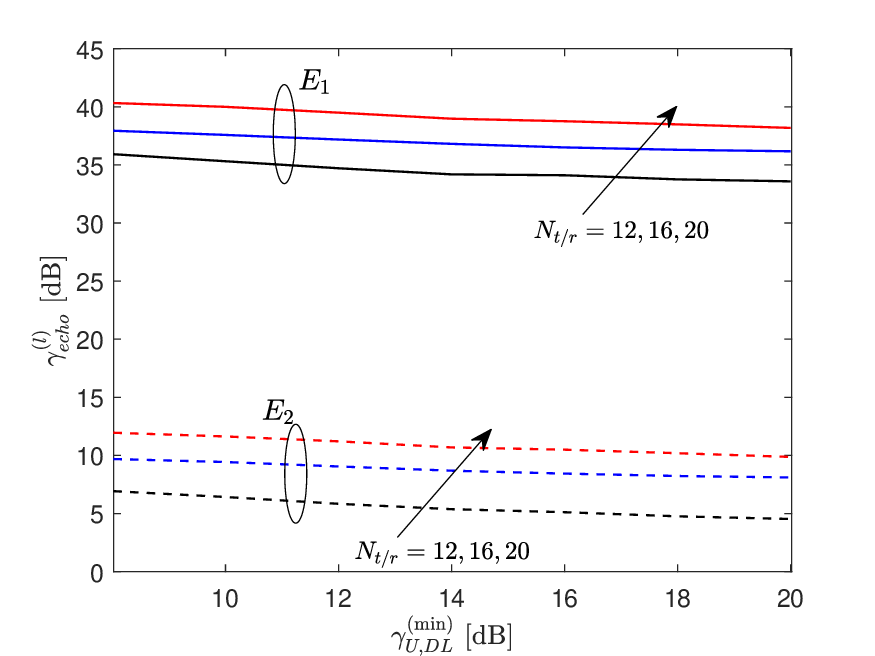}
\end{center}
\par
\vspace*{-.2cm}
\caption{$\protect\gamma _{\mathrm{echo}}^{(l)}$ of the $L$ targets in the
network vs. $\protect\gamma _{U,DL}^{(\min )}$.}
\label{fig4}
\end{figure}

To gain more insights on the secrecy-sensing trade-off discussed in Fig. \ref%
{fig3}, we present in Fig. \ref{fig4} the echo SNR, evaluated by %
\eqref{snrecho1}, for the considered $N_t$ and $N_r$ values of Fig. \ref%
{fig3}. The Rician K-factor values are set as in the previous figure. One can observe that the increase in the
minimal DL SINR requirement slightly reduces the echo SNR of both considered targets,
corroborating the observations made from Fig. \ref{fig3}. Notice
from the proportionality between \eqref{beampattern2} and \eqref{senspw} that the
lower the RIS beampattern gain, the lower the received sensing power at
the targets' plane. Consequently, this reduces the target detection's echo
SNR. For instance, the echo SNR drops from by $3$ dB when increasing $%
\gamma_{U,DL}^{(\min)}$ from $8$ to $20$ dB, reaching a minimal value of $34$
dB for $E_1$ (nearest target) and $4$ dB for $E_2$ (farthest target) with $%
N_t=N_r=12$. On the other hand, the increase in the transmit and receive
arrays size raises the transmit and receive beamforming gains at the BS,
increasing the echo SNR. In particular, a $3$-dB echo SNR gain is observed
by increasing $N_{t/r}$ from $12$ to $16$ and from $16$ to $20$.

\begin{figure}[tbp]
\begin{center}
\vspace*{-.4cm}

\includegraphics[scale=.48]{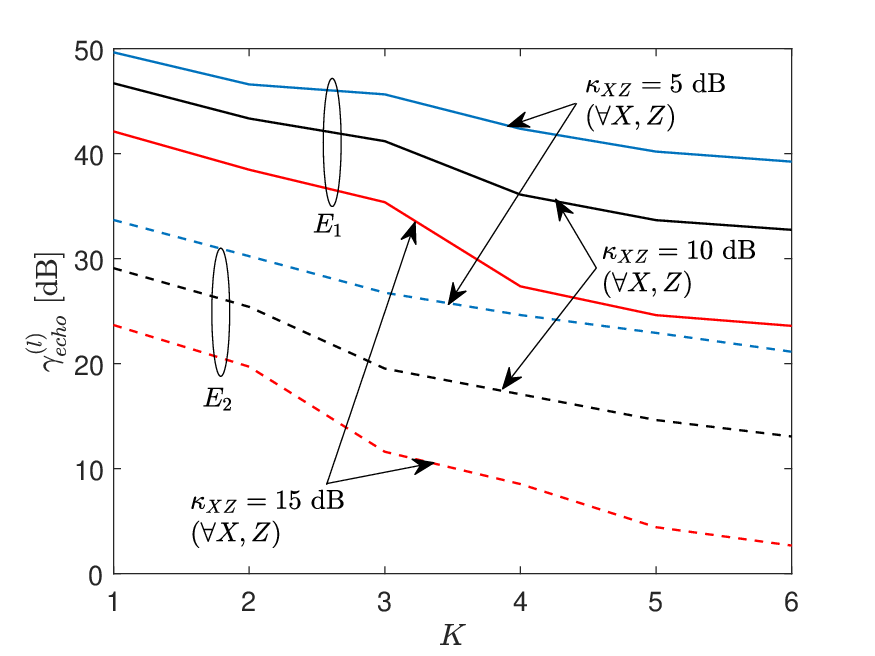}
\end{center}
\par
\vspace*{-.2cm}
\caption{$\protect\gamma _{\mathrm{echo}}^{(l)}$ of the $L$ targets in the
network vs. $K$.}
\label{fig6bs}
\end{figure}

In Fig. \ref{fig6bs}, the echo SNR for detecting the $L=2$ sensed targets is
shown as a function of the number of UL users $(K)$. In particular, the
sensing performance is shown for three different values of the various
links' Rician K-factor, i.e., $\kappa_{XZ}$ $
\left(\forall XZ \in \left\{BR, RU_j, V_kR \right\}\right) $,
namely $\kappa_{XZ}=5$ dB, $\kappa_{XZ}=10$ dB, and $\kappa_{XZ}=15$ dB. One
can observe that the echo SNR for both targets decreases with the increase
in the number of UL users, despite enhancing the overall network's UL sum SC
via allowing more users to access the network. In particular, notice an echo
SNR drop of $10$ dB approximately for both targets for $\kappa_{XZ}=5$ dB
when increasing $K$ from $1$ to $6$, whereby the echo SNR is equals $50$ dB
and $39$ dB for the $E_1$ and $34$, $22$ dB for $E_2$ for the $K=1$ and $K=6$, respectively. This can be explained by the fact that the incorporation of additional UL users increases the denominator of \eqref{snrecho1}, resulting
in a lower echo SNR. Furthermore, the inclusion of additional UL users
alters the optimized RIS configuration accordingly to fulfill their secrecy
requirements, affecting the maximal sensing performance. Thus, this presents
a sensing-UL sum SC trade-off in the system.  On the other hand, the
increase of the Rician K-factor reduces the echo SNR where, for instance,
when $K=6$, a $7$- and $9$-dB decrease of $\gamma_{\mathrm{echo}}^{(1)}$ is
observed when increasing $\kappa_{XZ}$ from $5$ to $10$ dB and from $10$ to $%
15$ dB, respectively, whereby a slightly higher drop is observed for $%
\gamma_{\mathrm{echo}}^{(2)}$. This is due to the fact that the decrease of $%
\kappa_{XZ}$ increases the contribution of the multipath components in the
different system's links, yielding a higher sensing signal power reaching
each target and, consequently, a higher echo SNR.

\begin{figure}[tbp]
\begin{center}
\includegraphics[scale=.48]{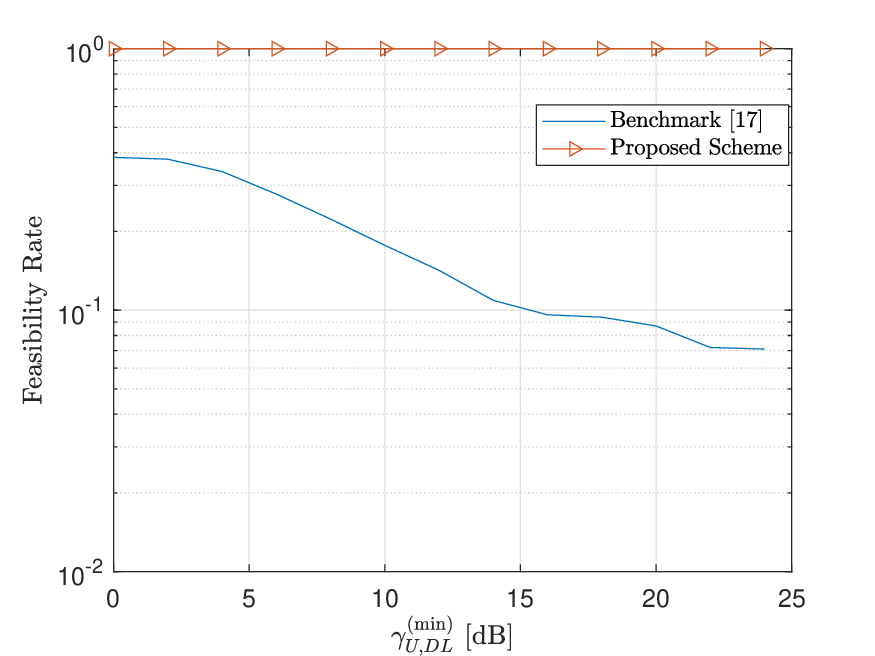}
\end{center}
\par
\vspace*{-.2cm}
\caption{Feasibility rate of the proposed algorithm vs. $\protect\gamma_{U,%
\mathrm{DL}}^{(\min)}$ compared with the benchmark one in \protect\cite%
{bazzi}.}
\label{figfeas}
\end{figure}

In Fig. \ref{figfeas}, the feasibility rate of the proposed AO-based
framework, given in Algorithm \ref{algg}, is shown versus $\gamma _{U,%
\mathrm{DL}}^{(\min )}$ and compared with the one proposed in the benchmark
FD-DFRC scheme in \cite{bazzi}. The feasibility rate represents the
percentage of times Algorithm \ref{algg} solved \eqref{P4m} and \eqref{P6m} successfully, evaluated by counting the number of times the convex programming
solver achieves a feasible solution divided by the number of simulations. For a fair and meaningful comparison, we
assume that all direct links between the different nodes in
the network are present for both schemes. For the proposed scheme, the AO-based framework in
Algorithm \ref{algg} was adapted by including a direct link component for
all the channels where similar SDP expressions for the various SINR and
sensing constraints were obtained. Furthermore, we set $J=L=K=1$, with $%
\varphi _{BZ_{1}}=10$ deg $(Z\in \{U,E\})$. Such a configuration represents
the scenario mentioned in \textit{Remark \ref{rkdirectlink}}, presenting a
case of insecure transmission for the baseline scheme in \cite{bazzi} when the eavesdroppers and DL users are
aligned with the same azimuth angle with respect to the BS. Also, we set $%
\varphi _{V_{1}B}=70$ deg and $\gamma _{E,\mathrm{DL}}^{(\max )}=\gamma _{E,%
\mathrm{UL}}^{(\max )}=0$ dB. One can observe that the benchmark scheme
yields a decreasing feasibility rate with the increase in the minimal
downlink SINR requirement. In fact, as $U_{1}$ and $E_{1}$ are aligned with
respect to $B$, the legitimate signal and AN are beamsteered to both users
due to their angular indistinguishability. Thus, it is expected that, in the
presence of a high LoS power (i.e., $\kappa_{XZ}=20$ dB), both received SINRs will be
of the same order. The increase (decrease) of $\gamma _{U,\mathrm{DL}%
}^{(\min )}$ ($\gamma _{E,\mathrm{DL}}^{(\max )}$) aims at forcing higher
(lower) $\gamma _{U_{1}}$ ($\gamma _{E_{1}}$) levels. Therefore, the higher $%
\gamma _{U,\mathrm{DL}}^{(\min )}$, the more unlikely the fulfillment of $%
\gamma _{U_{1}}>\gamma _{E_{1}}$ due to the superposition of the legitimate
signal with the AN, affecting the decoding of both $U_{1}$ and $E_{1}$
similarly. On the other hand, the proposed scheme maintains a $100\%$
feasibility rate in such a scenario, thanks to the RIS-aided beamforming,
whereby the RIS observes $U_{1}$ and $E_{1}$ in distinguished angles of
reflection, yielding a separate beamsteering of the legitimate signal and
the AN.

\section{Conclusion}

In this paper, the secrecy-sensing analysis and optimization of an
FD-DFRC-RIS network is carried out. By accurately modeling the various
secrecy and sensing performance metrics, represented by the SINRs, SCs,
sensing beampattern, and received sensing power at the targets, an
optimization problem for maximizing the network's sensing performance,
subject to secrecy and power constraints is formulated. By developing a
robust and rapidly converging AO-based scheme for solving the optimization
problem in hand, a maximal sensing performance is achieved while
guaranteeing the preset secrecy and power requirements. The results
manifest an existing secrecy-sensing trade-off, whereby the increase in
the number of UL users or in the minimal legitimate DL SINR
requirements degrades the sensing performance in terms of its echo SNR.
Furthermore, the proposed scheme outperforms its benchmark RIS-less FD-DFRC
one in the absence of a direct link or when the legitimate users are aligned
with the malicious targets.

\bibliographystyle{IEEEtran}
\bibliography{refs}

\end{document}